\documentclass[fleqn,usenatbib]{mnras}

\usepackage{newtxtext,newtxmath}

\usepackage[T1]{fontenc}

\DeclareRobustCommand{\VAN}[3]{#2}
\let\VANthebibliography\thebibliography
\def\thebibliography{\DeclareRobustCommand{\VAN}[3]{##3}\VANthebibliography}

\usepackage[pdftex]{graphicx}
\usepackage{amsmath}	
\usepackage{cleveref}
\crefformat{figure}{Fig.~#2#1#3}
\crefformat{table}{Table~#2#1#3}
\usepackage{natbib}
\usepackage{multirow}
\usepackage{xcolor}

\title[Mid-IR Variability of T Tauri Stars]{JWST-DECO: Temporal Variations in the 
Mid-IR Silicate Features of Two T Tauri Discs Based on Spitzer and JWST 
observations}

\author[N. Sameshima et al.]{
N. Sameshima,$^{1,2}$\thanks{E-mail: naotosame7010@g.ecc.u-tokyo.ac.jp}
T. Miyata,$^{3}$
T. Kamizuka,$^{3}$
Y. Aikawa,$^{1}$
M. Honda,$^{4}$
L. I. Cleeves,$^{5}$
N. P. Ballering,$^{6,7}$ 
\newauthor
M. J. Colmenares,$^{8}$
C. Gonz\'alez-Ruilova,$^{9,10,11}$
V. V. Guzman,$^{10,12}$
T. J. Haworth,$^{13}$
C. J. Law$^{5,14}$
\newauthor
and
J. P. Williams$^{15}$
\\
$^{1}$Department of Astronomy, Graduate School of Science, The University of Tokyo, 7-3-1 Hongo, Bunkyo-ku, Tokyo 113-0033, Japan\\
$^{2}$Japan Aerospace Exploration Agency, Institute of Space and Astronautical Science, 3-1-1 Yoshinodai, Chuo-ku, Sagamihara-shi, Kanagawa
252-5210, Japan\\
$^{3}$Institute of Astronomy, Graduate School of Science, The University of Tokyo, Osawa 2-21-1, Mitaka, Tokyo 181-0015, Japan\\
$^{4}$Department of Biosphere-Geosphere Science, Okayama University of Science, 1-1 Ridai-cho, Kita-ku, Okayama 700-0005, Japan\\
$^{5}$Department of Astronomy, University of Virginia, Charlottesville, VA 22904, USA\\
$^{6}$Space Science Institute, Boulder, CO 80301, USA\\
$^{7}$Department of Astronomy, University of Wisconsin-Madison, Madison, WI 53706, USA\\
$^{8}$Department of Astronomy, University of Michigan, Ann Arbor, MI 48109, USA\\
$^{9}$Departamento de F\'isica, Universidad de Santiago de Chile, Av. Victor Jara 3659, Santiago, Chile\\
$^{10}$Millennium Nucleus on Young Exoplanets and their Moons (YEMS), Av. Victor Jara 3659, Santiago, Chile\\
$^{11}$Center for Interdisciplinary Research in Astrophysics and Space Exploration (CIRAS), Universidad de Santiago de Chile, Av. Victor Jara 3659, Santiago, Chile\\
$^{12}$Instituto de Astrof\'isica, Pontificia Universidad Cat\'olica de Chile, Av.\ Vicu\~na Mackenna 4860, 7820436 Macul, Santiago, Chile\\
$^{13}$Astronomy Unit, Department of Physics and Astronomy, Queen Mary University of London, Mile End Road, London E1 4NS, UK\\
$^{14}$NASA Hubble Fellowship Program Sagan Fellow\\
$^{15}$Institute for Astronomy, University of Hawaii, Honolulu, HI 96822, USA
}

\date{Accepted XXX. Received YYY; in original form ZZZ}

\pubyear{\the\year{}}

\begin{document}
\label{firstpage}
\pagerange{\pageref{firstpage}--\pageref{lastpage}}
\maketitle

\begin{abstract}
	Mid-infrared spectra of planet-forming discs commonly show 
	prominent silicate emission, whose spectral shape is sensitive 
	to the disc temperature distribution as well as its 
	mineralogical composition. We report new James Webb Space Telescope 
	(JWST) observations of the discs around Sz 96 and IP Tau
  and find that their silicate emission significantly 
	changes in the 20 years since they were observed with the 
	Spitzer Space Telescope (SST). 
	Significant differences between the SST and JWST spectra are  
	found for both sources, with flux variations of  
	10--15\% in Sz~96 and 30--35\% in IP Tau. Sz 96 dimmed at $\le$ 18~$\mathrm{\mu m}$
	and did not change significantly at longer wavelengths,
	whereas IP Tau became brighter across the entire wavelength range, 
	with a particularly strong enhancement around 10~$\mathrm{\mu m}$ in 
	the JWST data compared to the SST data. We propose that this large degree 
	of variability is explained by structural changes in the inner regions of the discs.
	Specifically, we also find that crystalline silicates exhibit lower temperatures 
	than amorphous silicates in the JWST data of both sources. This result 
	supports the idea that crystalline grains, formed through high-temperature 
	annealing in the inner disc regions, have been transported outward, 
	leading to their presence in cooler regions of the disc.
	While similar behavior had been reported in previous SST-based studies, 
	the much higher spectral resolution of JWST 
	enables clearer identification of the crystalline features.
\end{abstract}




\section{Introduction}\label{s1}
T Tauri stars host discs, which are the 
sites of planet formation. 
By studying their structure, dust composition, and 
dynamics, we can gain insights into  
the process of planet formation and disc evolution  
\citep{W2011}.

The disc is heated by radiation from the central star 
and emits over a wide range of wavelengths.
Recent ALMA observations at 1.3 mm trace cold dust 
located at distances of a few to several hundred au, 
revealing various substructures such as rings and cavities \citep{O2025}.
Near- and mid-infrared emission originates from warmer regions 
located at distances of approximately 0.1 to a few au \citep{D2010}. 
According to the infrared excess relative 
to a stellar photosphere, the discs are classified into three categories  
\citep{E2014}: (1) full disc, where the infrared excess  
consists of dust emission with a range of temperatures,  
suggesting that the disc is continuous from the inner  
region to the outer region; (2) pre-transitional disc,  
which has a gap separating the inner and outer discs;  
(3) transitional disc, which has a central hole, i.e.  
an almost complete absence of dust emission in the  
inner region of the disc. 

Regarding the dust composition, silicates are the main  
components, and their mid-infrared emission bands arise  
from the hot surface layers of the discs. In addition to the  
characteristic amorphous silicate features around  
10~$\mathrm{\mu m}$, crystalline silicate features have also 
been detected (\citealt{H2003}; \citealt{B2005};  
\citealt{K2006}; \citealt{W2008}). However, in the  
interstellar medium, silicate dust is predominantly  
observed in an amorphous state \citep{K2004}. This suggests that  
some degree of silicate crystallisation processes must be  
taking place in protoplanetary discs.

While the exact crystallisation mechanism remains a topic of  
debate, one of the leading hypotheses is thermal  
annealing in the high-temperature inner disc region, typically 
at $\sim$ 1000 K and $\sim$ 0.1 au \citep{B2002}. However, there have  
been reports of crystalline features even in the outer  
regions of the discs, located at a few au from the star, 
where the temperature is too low for  
annealing \citep{W2008}. \citet{O2009} reported that  
around T Tauri stars, crystalline features are detected  
about 3.5 times more frequently in the colder outer  
regions than in the warmer inner regions. \citet{O2010} pointed out  
that the higher detection rate of crystalline features  
in the colder component does not necessarily mean their  
higher abundance in the outer region. It could instead  
result from a contrast problem, where the strong 
10~$\mathrm{\mu m}$ feature of amorphous silicates in the  
warmer inner disc masks the weaker crystalline features.  
To resolve this contrast problem and reliably  
determine the distribution of crystalline silicates,  
we need higher resolution data with high signal-to-noise ratio, 
which is now available thanks to the James Webb Space Telescope (JWST;  
\citealt{R2023a}).

In addition to the dust composition, temporal variability  
of the spectra provides key insights into disc evolution.  
Young stellar objects (YSOs), including T Tauri stars,  
exhibit variability on time scales of days to years in  
both optical and infrared wavelengths (\citealt{S2014};
\citealt{B2014}; \citealt{M2011}; \citealt{S2014a}; \citealt{W2010};
\citealt{M2011a}). These observations have revealed the 
diverse nature of YSO variability.
Various mechanisms contribute to the optical variability, including  
long-term increases in brightness due to disc instabilities  
(\citealt{H1996}), short-term brightness enhancements caused  
by magnetically induced instabilities (\citealt{A2010}), and  
periodic variations tracing the rotation of magnetic spots  
on the stellar surface (\citealt{G2008}). Mid-infrared spectral variation  
can provide information about the dust composition and temperature 
distribution of discs. The spectra obtained by the Infrared  
Spectrograph (IRS; \citealt{H2004}) on-board the Spitzer 
Space Telescope (SST; \citealt{W2004}) have  
been especially useful. \cite{E2011} examined the SST IRS  
spectra of pre-transitional discs taken between 2005 and  
2008 to find temporal variations, which can be explained  
by 10--20\% change of the inner disc scale height. Temporal  
variability of a disc spectrum over $\sim 20$ years can  
now be investigated using SST and JWST data. 
This offers insights into dynamical changes occurring within, at most, 
the inner few au, although the observed variability is 
more likely to trace regions much closer to the inner rim.
\cite{J2024} investigated spectral variations in the T Tauri star  
PDS~70 by comparing the SST IRS data and the JWST  
Mid-Infrared Instrument Medium Resolution Spectroscopy  
(MIRI MRS; \citealt{A2023}) data, revealing changes in  
temperature distribution and dust composition. However, the  
number of studies comparing the SST and JWST spectral  
data remains limited. It is important to increase the  
sample to understand the evolution of physical structure  
and dust composition in discs, as the current sample size 
is too small to draw statistically meaningful conclusions. 

In this study, we investigate the variability of two  
T Tauri stars that have been observed with both the SST  
and the JWST. By performing model fitting of the  
mid-infrared spectral data, we quantitatively evaluate  
the changes in the disc's temperature structure and dust  
composition. In addition, the JWST data have higher 
spectral resolution and signal-to-noise ratio than the 
SST data. This enables clearer identification of 
silicate features and provides important clues about 
the distribution of crystalline dust, which was previously 
unclear. In \Cref{s2}, we describe the observations and data  
processing, while \Cref{s3} outlines the model fitting  
methodology. The results are presented in \Cref{s4},  
followed by a discussion in \Cref{s5}, and conclusions  
in \Cref{s6}.

\section{Observations and Data}\label{s2}
\subsection{Targets}\label{s2-1}
For this study, we selected two T Tauri stars, Sz 96  
and IP Tau, as our targets. They are among 80 targets  
of the Disc-Exoplanet C/Onnection (DECO) ALMA Large 
Program (PI: Ilse Cleeves; 2022.1.00875.L) to investigate 
the gas composition in a large sample of protoplanetary discs.
The JWST spectra were taken in its follow-up programme, JWST-DECO  
(Proposal ID: 3228; PI: Ilse Cleeves). 
We selected Sz 96 and IP Tau based on 
three criteria: the availability of SST spectroscopic data 
over a wide wavelength range, the absence of nearby bright 
objects, and the presence of strong 10~$\mu$m silicate features.
The basic properties of these stars are summarised in \cref{t01}.  
Their stellar properties are similar, but their disc structures 
are different, as described in the following.

\textit{Sz 96}: \cite{B2020} investigated the colour index  
of dipper stars in the Lupus star-forming region using  
2MASS $K_s$ band, WISE W3 (12~$\mathrm{\mu m}$) and W4  
(22~$\mathrm{\mu m}$) bands and showed that the observed  
infrared excess of this object is consistent with that  
of a full disc. In the millimetre wavelength range observed
by ALMA, \cite{G2025} showed the dust mass of 0.5~$M_\oplus$ and 
identified a ring with a radius of about 6~au, which
corresponds to a colder outer region than discussed by \cite{B2020}.
Since inner warm dust with a mass comparable to that of 
the outer ring would produce detectable millimetre emission, the 
observation indicates that the mass of warm, mid-infrared-emitting 
dust in the inner disc is much smaller than that in the outer ring.
\cite{X2023} independently estimated the disc radius 
to be 41 au and the dust mass to be 0.74~$M_\oplus$.
\cite{O2010} modelled the SST IRS  
spectrum of Sz 96 using a combination of five dust  
components -- olivine, pyroxene, and silica for the  
amorphous phase, and forsterite and enstatite for the  
crystalline phase -- with a warm component at 230\,K and a  
cold component at 90\,K.   
\cite{B2020} detected variability in the $V$ band based  
on photometric observations by ASAS-SN from 2016 to 2019  
and classified this star as a quasi-periodic dipper.  
\cite{P2021} reported that Sz~96 showed variability  
of up to 0.35 mag in the WISE W2 band (4.6~$\mathrm{\mu m}$)  
over a 6.5-year period since 2013, indicating that this object
is intrinsically variable in the mid-infrared.

\textit{IP Tau}: Based on an analysis of the spectral  
index and the equivalent width of the 10~$\mathrm{\mu m}$  
silicate feature, \cite{F2009} classified the disc of IP  
Tau as a pre-transitional disc. \cite{E2011} modelled the  
SST IRS data for IP Tau, proposing an inner and outer disc 
structure separated by a gap lacking optically thick material, 
with inner-edge temperatures of 1400 and 230\,K, respectively. A  
high-resolution observation by ALMA revealed a ring-like  
structure with a 10~au-wide ring at a radius of  
27~au \citep{L2018}, which is much farther out than the  
structures traced by mid-infrared observations. According  
to \cite{P2021}, IP Tau exhibited variability of up to  
0.37~mag in the W2 band during NEOWISE photometric  
monitoring over a 6.5-year period starting in 2013. IP  
Tau was observed multiple times by SST, and a variation  
in the continuum slope was detected. The variation was explained by  
a change in the height of the inner wall by \cite{E2011}.

\begin{table}
	\centering
    \caption{Stellar properties of the targets.}
    \label{t01}
    \begin{tabular}{lcc}
        \hline
        Property & Sz 96 & IP Tau \\
        \hline\hline
        Spectral type & M1$\pm$0.5 & K7 \\
        $T_\mathrm{eff}$ (K) & 3705$\pm$171 & 3770$\pm$150 \\
        Distance (pc) & 156$\pm$4 & 129 \\
				Inclination (deg) & -- & 45 \\
        Stellar Radius ($\mathrm{R}_\odot$) & 2.0$\pm$0.5 & 1.3 \\
        Stellar Mass ($\mathrm{M}_\odot$) & 0.5$\pm$0.1 & 0.51 \\
        Disc type & Full & Pre-Transitional \\
        \hline
    \end{tabular}
	\begin{minipage}{\linewidth}
			\centering
			\small
			\parbox{\linewidth}{
			References: For Sz 96, the spectral type, distance, 
			radius and mass are from \cite{X2023}, and
			the effective temperature is from \cite{A2017}, while the 
			inclination is not available. For IP Tau, the inclination
			is from \cite{L2018} and the other properties are from \cite{G2022},
			with uncertainties available only for the effective temperature.}
	\end{minipage}
\end{table}

\subsection{SST IRS}\label{s2-2}
Sz 96 was observed with the SST IRS on 15 March 2006  
(programme ID: 179; PI: N. Evans), and IP Tau was  
observed on 8 October 2008 (programme ID: 50403; PI:  
N. Calvet). Both observations were carried out using  
the Short-Low (SL) and Long-Low (LL) modules of the IRS,  
covering a wavelength range of 5.2--38~$\mathrm{\mu m}$  
with a spectral resolution of $R=$60--130 ($\Delta v=$ 
2300--5000 $\mathrm{km\,s^{-1}}$).

The spectroscopic data used in this study were taken from  
the IRS Enhanced Spectrophotometric Catalogue available  
on the Spitzer Heritage Archive\footnote{\url{https://irsa.ipac.caltech.edu/applications/Spitzer/SHA/}}
(SHA; \citealt{W2010a}).
In the reduction process of the data, the one-dimensional  
spectra were taken from the background-subtracted  
Post-BCD ("\texttt{bksub.tbl}") products. For each nod  
position, the median spectrum was computed from all  
exposures, and the two nod spectra were then averaged to  
produce the final spectrum. The flux uncertainty for each  
nod spectrum was derived from the "\texttt{bkunc.fits}"  
file, and the final uncertainty was computed by adding  
the two in quadrature \citep{I2021}.

We compared the spectra with those obtained from the 
Cornell Atlas of Spitzer/IRS Sources 
(CASSIS\footnote{\url{https://cassis.sirtf.com/}}; \citealt{L2015}), 
which provides publishable-quality, flux-calibrated IRS spectra, 
and found that the fluxes between the two datasets are consistent 
within the uncertainties. The weighted mean of $(F_\mathrm{CAS}/F_\mathrm{SHA} - 1)$ is
$2.87 \pm 0.17$\% for Sz~96 and $2.41 \pm 0.41$\% for IP~Tau,
indicating that the CASSIS spectra are slightly brighter than those from SHA.
However, the differences are small ($\le$~3\%) within the expected 
calibration uncertainties, and are not expected to
substantially alter our interpretation of the observed variability.

\subsection{JWST MIRI MRS}\label{s2-3}
Sz 96 and IP Tau were observed with JWST MIRI MRS on  
28 July 2024, as part of the GO Cycle 2 JWST-DECO programme.  
The wavelength range covered by the MRS mode is  
4.9--27.9~$\mathrm{\mu m}$, with a spectral resolution of  
$R=$ 1500--3500 ($\Delta v=$ 100--200 $\mathrm{km\,s^{-1}}$). 

The spectral data were processed using the JWST
Calibration Pipeline v1.18.0 \citep{B2025} and further
reduced with the JDISCS MIRI MRS pipeline v9.1 developed 
for the JWST JDISCS survey \citep{A2025}.

We binned the spectra to a resolution of $R = 500$ ($\Delta v=$ 
600~$\mathrm{km\,s^{-1}}$), which is still much higher than that of SST IRS, to avoid 
issues with the Non-Negative Least-Squares (NNLS) calculation 
during the model fitting (as described in \Cref{s3-3}) due 
to the high spectral resolution. 
This resolution is sufficiently high to clearly capture  
the dust features described in \Cref{s3-2}, and was also  
adopted by \citet{J2024}, who studied the variability in
PDS 70 based on the SST and JWST observations.

\subsection{Spectra}\label{s2-4}
\begin{figure}
    \centering
    \includegraphics[width=\linewidth]{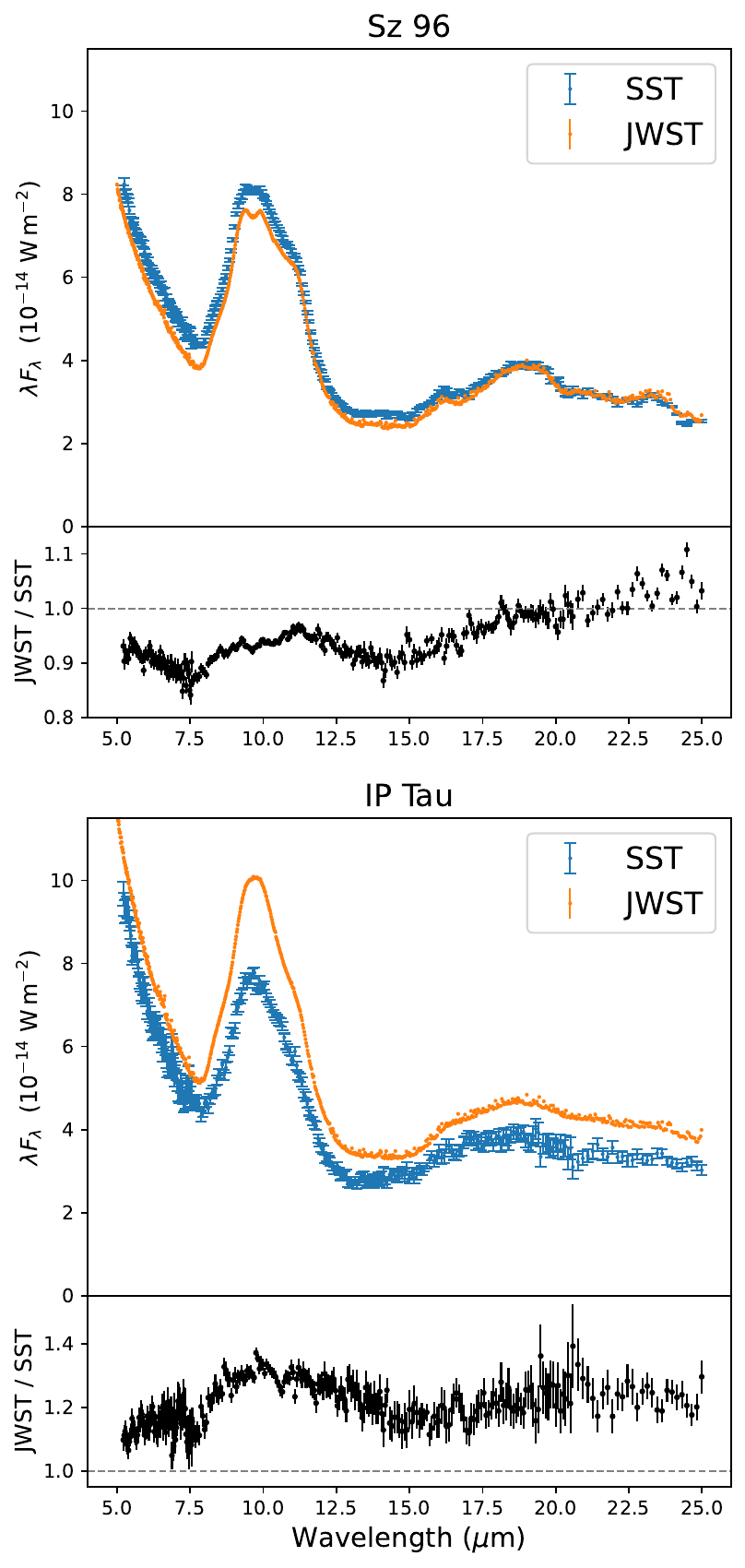}
    \caption{The SST (capped error bars) and JWST (vertical error bars) spectra and their ratio 
		(JWST/SST) for Sz 96 (top) and IP Tau (bottom).
    The JWST data have been binned to a resolution of $R=500$.}\label{f01}
\end{figure}

The spectral data are shown in \cref{f01}. In the  
following sections, we focus on the 5--25~$\mathrm{\mu m}$  
range to compare the SST and JWST data. In the range of  
5--7.5~$\mathrm{\mu m}$, radiation from high-temperature  
optically thick components including the central star is  
observed, while amorphous silicate features from  
optically thin components are seen at longer wavelengths  
(e.g. the peak around 10~$\mathrm{\mu m}$ and the bumps  
at 18~$\mathrm{\mu m}$). At longer
wavelengths, sub-peaks characteristic of crystalline 
silicates can also be seen. These features have 
also been observed in other T~Tauri stars (e.g. \citealt{F2006}).

For both objects, the overall spectral shapes are  
similar between the SST and JWST spectra, but a temporal  
variation is clearly found. Sz 96 shows dimming at 
$\leq$18~$\mathrm{\mu m}$, while showing almost no variation 
at longer wavelengths. IP~Tau brightens across the entire wavelengths, 
with a particularly strong enhancement around 10~$\mathrm{\mu m}$.
The flux differences reach 10--15\% for Sz~96 and 30--35\% for IP Tau.

\section{Fitting}\label{s3}
\begin{figure}
	\centering
	\includegraphics[width=0.8\linewidth]{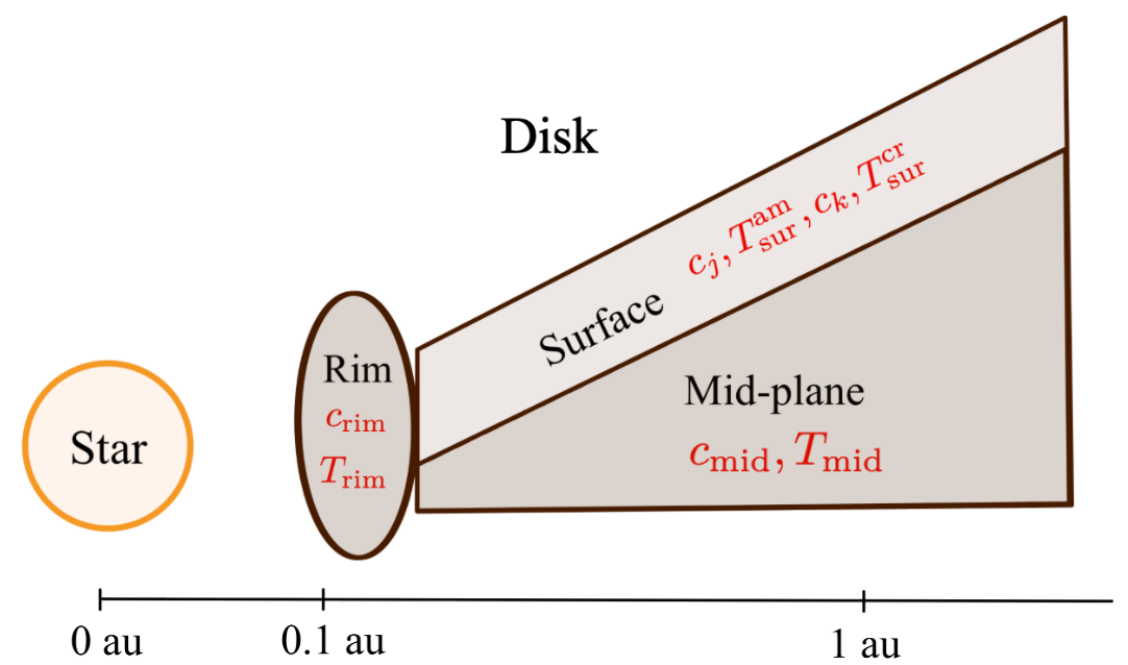}
	\caption{An illustration of the model used in our study.
	The star, rim and mid-plane are optically thick and the 
	surface is optically thin. Silicate features are seen in 
	the radiation from the surface.}\label{f02}
\end{figure}

To investigate the temperature structure and dust  
composition of the discs, we performed model fitting to  
the spectral data. We used a model based on the Dust Continuum Kit (DuCK;  
\citealt{K2024}), which was also employed in the  
analysis of SST and JWST spectral data for PDS~70 by  
\cite{J2024}. The model assumes contributions from  
several components: the central star, a vertically  
extended hot inner rim, the relatively cooler mid-plane  
of the disc, and the surface layer (see \cref{f02}).  
While the inner rim and disc mid-plane are optically  
thick, the surface layer is optically thin. The emission  
from the optically thin surface produces the silicate  
features, and the model includes multiple species of  
silicate dust (see \Cref{s3-2}). The fitting parameters include the  
temperature, solid angle, and dust mass of each  
contributing component. In the original DuCK model, the  
surface layer and mid-plane have a radial temperature  
distribution. In the present study, however, we assume  
each component has a single temperature for simplicity. Although we also  
tried a model with radial temperature distributions, the  
quality of the fit did not significantly change. 
This minor modification does not affect the conclusions presented later in the paper.
We performed the parameter fitting using our own Markov Chain 
Monte Carlo (MCMC) implementation, rather than the built-in 
nested sampling routine provided by DuCK.

\subsection{Model}\label{s3-1}
The model spectrum is expressed as follows:
\begin{align}
    F_\mathrm{model}=F_\mathrm{star}+F_\mathrm{rim}+
    F_\mathrm{mid}+F_\mathrm{sur},
\end{align}
where $F_\mathrm{star}$ is the emission from the star,  
$F_\mathrm{rim}$ is from the inner rim, $F_\mathrm{mid}$  
is from the mid-plane of the disc, and $F_\mathrm{sur}$  
is from the optically thin surface layer.

The emission from the star is given by:
\begin{align}
    F_\mathrm{star} = \frac{\pi R^2_\mathrm{star}}{d^2}B(T_\mathrm{star}),
\end{align}
where $d$ is the distance to the star, $R_\mathrm{star}$ is the  
stellar radius, $T_\mathrm{star}$ is the stellar  
temperature, and $B(T)$ represents the Planck function  
with temperature $T$. We adopted the literature values  
listed in \cref{t01} for these parameters, and  
$F_\mathrm{star}$ is fixed in the model calculation.

The emission from the optically thick rim and mid-plane  
components is similarly expressed as:
\begin{align}
    &F_\mathrm{rim}=c_\mathrm{rim}B(T_\mathrm{rim}),\\
    &F_\mathrm{mid}=c_\mathrm{mid}B(T_\mathrm{mid}),
\end{align}
where $T_\mathrm{rim}$ and $T_\mathrm{mid}$ are the  
temperatures of the rim and mid-plane, respectively. The  
parameters $c_\mathrm{rim}$ and $c_\mathrm{mid}$  
correspond to the solid angles subtended by each  
radiation source as seen by the observer.

The emission from the optically thin surface layer is  
modelled by allowing two temperatures: one for amorphous  
grains ($T^\mathrm{am}_\mathrm{sur}$) and one for  
crystalline grains ($T^\mathrm{cr}_\mathrm{sur}$). The  
total surface emission is then given by:
\begin{align}
    F_\mathrm{sur}=\sum_{j} c_j \kappa_j B(T^\mathrm{am}_\mathrm{sur}) + 
		\sum_{k}  c_k \kappa_k B(T^\mathrm{cr}_\mathrm{sur}),
\end{align}
where $j$ and $k$ represent the amorphous and crystalline  
silicate dust species included in the model, and $\kappa$  
is the mass absorption coefficient. We note that scattering 
is negligible in the optically thin surface layer \citep{J2024}.
In this formulation, $c$ is proportional to the dust mass present in the  
surface layer.

Although our model assumes a constant stellar flux, we note that 
both objects can show optical variability that may be 
related to dipper-like extinction events. However, since the 
extinction efficiency decreases rapidly toward the mid-infrared, 
for instance, \cite{W2019} report $A_\lambda/A_V = 0.026$ at 
$\lambda = 4.5\,\mu{\rm m}$, the corresponding flux change  
would be negligible. Therefore, the assumption is not expected 
to introduce significant uncertainty in our analysis.

\subsection{Dust mass absorption coefficient}\label{s3-2}

\begin{figure*}
	\centering
	\includegraphics[width=\linewidth]{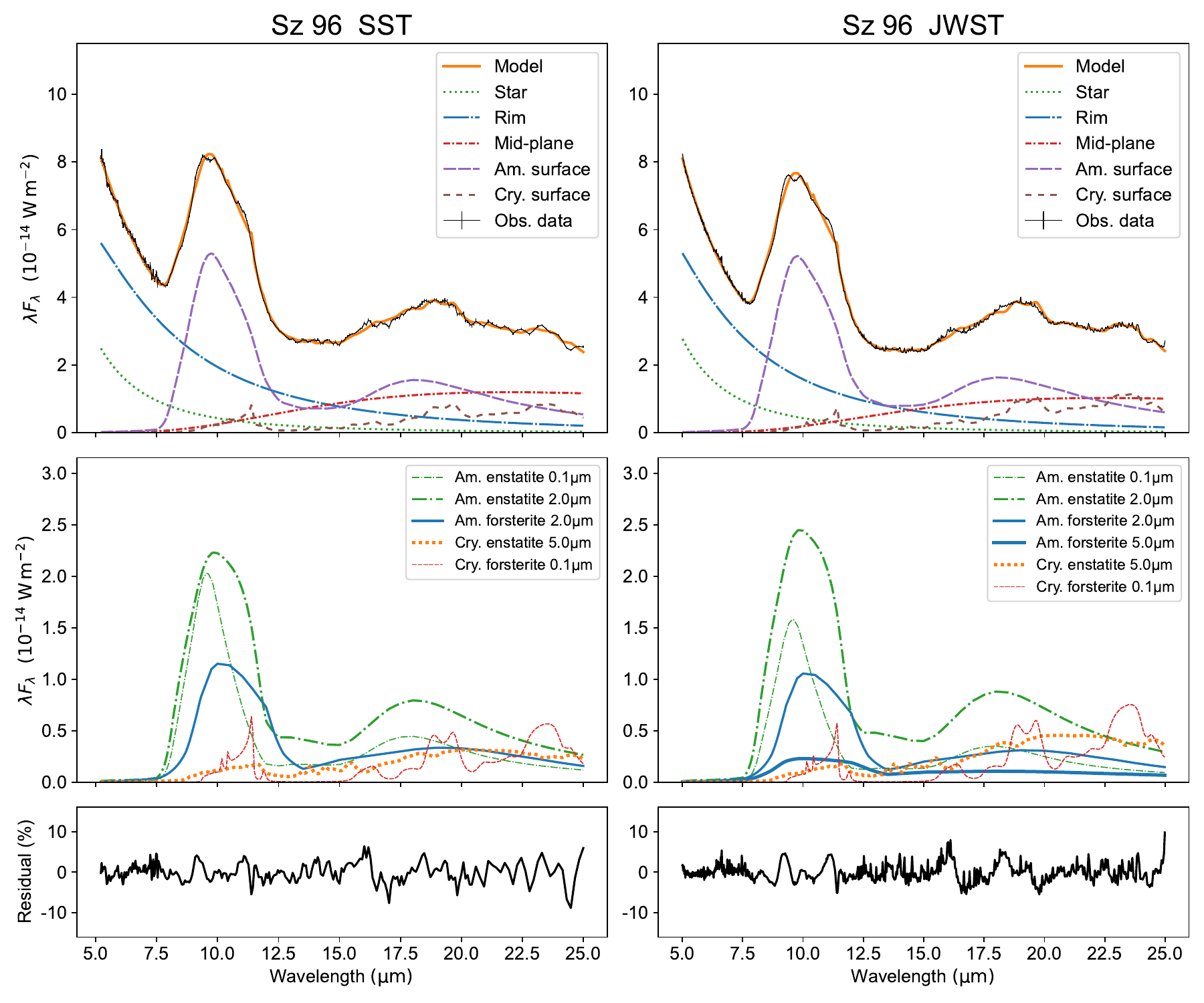}
	\caption{Fitting results of Sz 96: overall fitting results 
	(top), contributions from each dust component (middle), and 
	residual of the best-fitting model (bottom) for SST (left) and 
	JWST (right). In the middle panel, the emissions from the star, 
	rim, and mid-plane, as well as components contributing less than 
	0.5\% to the total flux, are not shown.}\label{f03}
\end{figure*}

\begin{figure*}
	\centering
	\includegraphics[width=\linewidth]{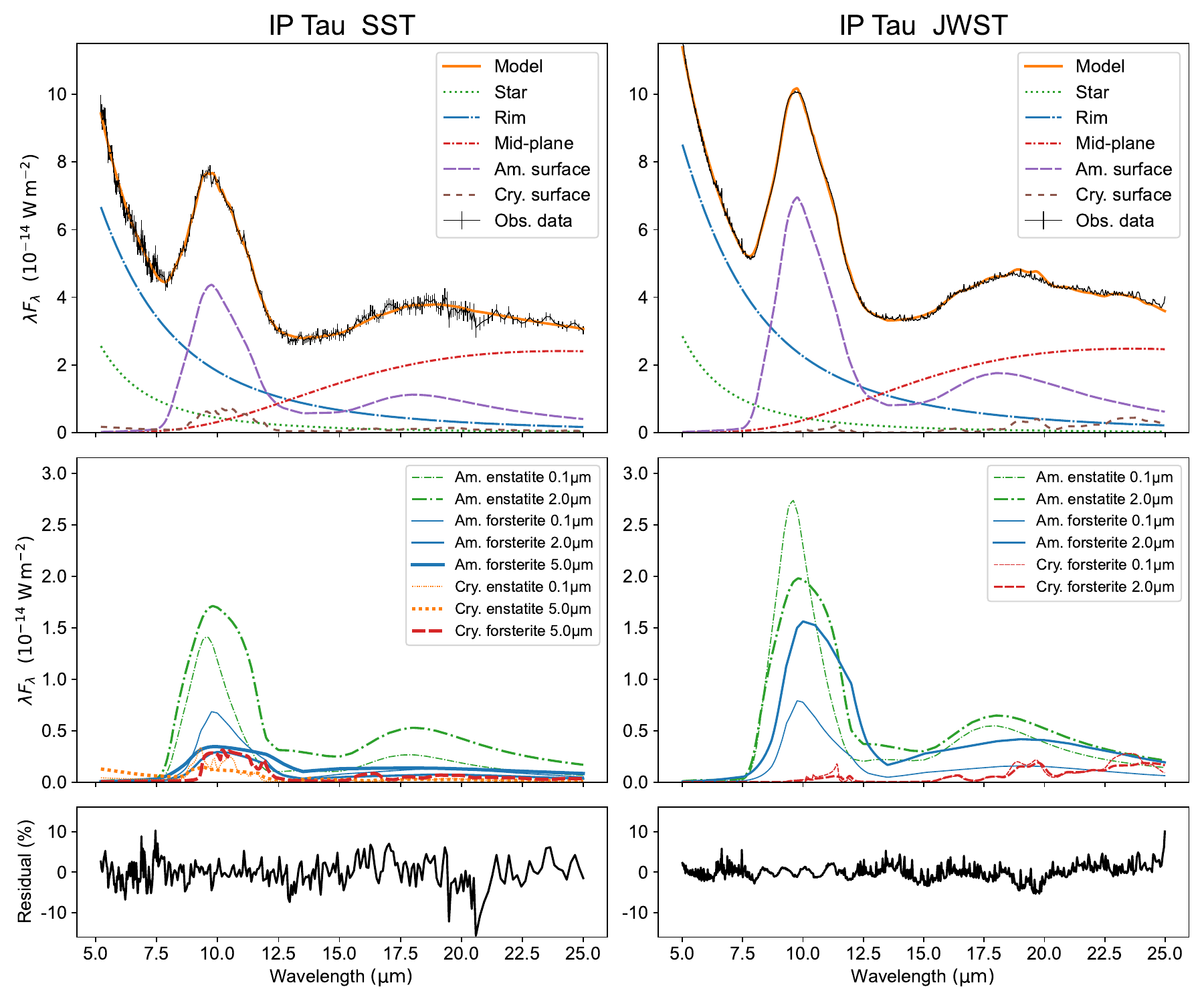}
	\caption{Fitting results of IP Tau: overall fitting results 
	(top), contributions from each dust component (middle), and 
	residual of the best-fitting model (bottom) for SST (left) and 
	JWST (right). In the middle panel, the emissions from the star, 
	rim, and mid-plane, as well as components contributing less than 
	0.5\% to the total flux, are not shown.}\label{f04}
\end{figure*}

We considered five dust species in the model following \cite{J2024}
: amorphous forsterite, amorphous enstatite, amorphous silica,  
crystalline forsterite, and crystalline enstatite. For 
each species, grain sizes of 0.1, 2, and 5~$\mathrm{\mu m}$ 
were adopted following \cite{J2010}. 
We tested 6 grain sizes (0.1, 1, 2, 3, 4, and 5~$\mathrm{\mu m}$) as 
in \cite{J2024}, but found that the fitting quality did not improve significantly.
Therefore, we adopted the 3 grain sizes for simplicity.
These sizes are typical in protoplanetary discs, and grain sizes greater than 5~$\mathrm{\mu m}$ 
are not considered because the band strength becomes very 
weak already for 5~$\mathrm{\mu m}$-sized grains \citep{J2024}.
The optical constants of each dust species were obtained from  
the literature listed in \cref{t02}. The mass absorption  
coefficients were calculated using the optical constants  
with the Python package \textit{Optool} (\citealt{D2021}).  
We employed the Distribution of Hollow Spheres (DHS;	
\citealt{M2003}) model in \textit{Optool}, which calculates the average  
optical properties of dust aggregates with complex shapes  
by approximating them as hollow spheres. This statistical 
approach avoids the artificial resonances associated with 
idealized spherical grains and provides a more realistic 
representation of the optical properties of irregular and 
porous dust particles in protoplanetary discs (\citealt{M2016}; \citealt{W2016}).
The parameter $f_{\text{max}}$, representing the hollow fraction, was  
set to $f_{\text{max}}=0.7$ for amorphous dust and  
$f_{\text{max}}=0.99$ for crystalline dust,
as these values are appropriate for reproducing 
the features of each dust type (\citealt{M2005}; \citealt{M2007}).
The optical properties of crystalline silicates depend on the relative 
orientation between the incident radiation field and the crystallographic axes.
Two approaches can be used to calculate the absorption coefficient 
taking into account the anisotropy of the refractive index. 
One is the Bruggeman rule, which treats each grain as an ensemble 
of sub-wavelength crystalline domains that are randomly oriented 
within the particle, thereby deriving an effective dielectric 
response for the grain as a whole \citep{B1935}.
The other is the three-axis average, which treats each grain 
as a single crystal and computes the absorption along the three 
principal crystallographic axes, then averages these values 
assuming a random orientation of the grains.
In this study, we adopted the Bruggeman rule, and the results are 
presented in \Cref{s4}. We also tested the simple averaging method, 
and the results are shown in Appendix~\ref{app:a}.

\begin{table}
    \centering
    \caption{The references of optical constants to calculate 
    the mass absorption coefficients.}\label{t02}
    \begin{tabular}{ll}
    \hline
    Species & Reference  \\
    \hline\hline
    Amorphous forsterite & \cite{J2003}  \\
    Amorphous enstatite & \cite{D1995} \\
    Amorphous silica & \cite{K2007} \\
    Crystalline forsterite & \cite{S2006} \\
    Crystalline enstatite & \cite{J1998} \\
    \hline
    \end{tabular}
\end{table}

\subsection{Fitting}\label{s3-3}
The fitting parameters are the temperatures $T$ and  
coefficients $c$ of each component, making a total of 21  
parameters. In this fitting, we combined the MCMC method  
with the NNLS method. First, MCMC proposes a set of four  
temperatures ($T_{\text{rim}}, T_{\text{mid}},  
T^\text{am}_\mathrm{sur}, \mathrm{and} \  
T^\text{cr}_\mathrm{sur}$). For the given temperatures,  
the set of $c$ values that best matches the observational  
data is computed by NNLS. Once the parameter set is  
determined, the model spectrum can be computed, and the  
likelihood, expressed by the following equation, can be  
calculated:
\begin{align}
    \mathcal{L} =\prod_{i=1}^{N_{\mathrm{obs}}} 
    \frac{1}{\sqrt{2\pi\sigma_i}}
    \exp\left[-\frac{(F_{\mathrm{model},i}-F_{\mathrm{obs},i})^2}
    {2\sigma_i^2}\right],
\end{align}
where $i$ represents each data point, and $\sigma$ is the  
uncertainty in the observational data. This allows the  
MCMC process to proceed to the next step using the  
updated temperature parameters. We set the prior ranges  
for the temperature of the rim, mid-plane, amorphous and  
crystalline dust in the surface to be  
$T_\mathrm{rim}=[500,1900]\ \mathrm{K},T_\mathrm{mid}=[10,700]\ \mathrm{K},  
T^\text{am}_\mathrm{sur}=[10,1000]\ \mathrm{K}, \mathrm{and} \  
T^\text{cr}_\mathrm{sur}=[10,1500]\ \mathrm{K}$, respectively.  
For the MCMC method, we used the Python package \textit{emcee}  
(\citealt{F2013}), and for the NNLS method, we used the  
nnls function in the \textit{SciPy} (\citealt{V2020}) library. In  
\textit{emcee}, the number of walkers and steps were set to 32 and 5000,
respectively, and the default StretchMove with the stretch scale parameter  
$a = 2.0$ was used (\citealt{G2010}).
The fitting was performed independently for the SST and JWST data.

\section{Results}\label{s4}

\begin{figure}
			 \centering
			 \includegraphics[width=\linewidth]{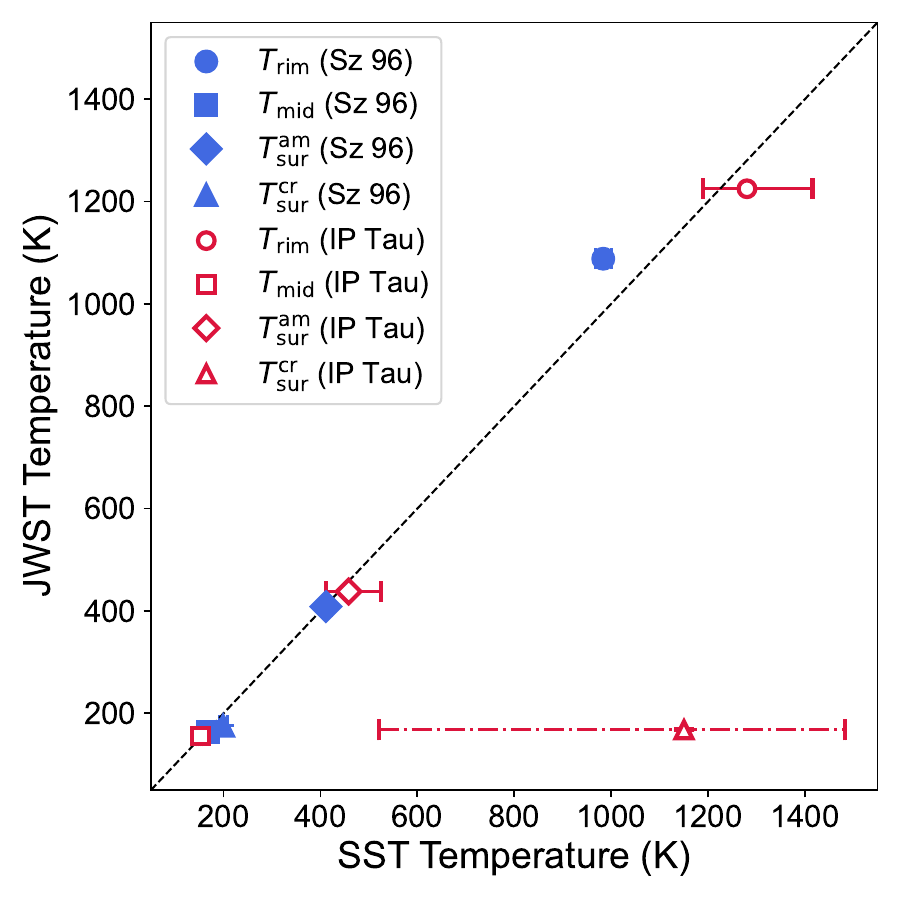}
	 \caption{The temperature parameters for Sz 96 (filled symbols) and IP Tau (open
	 symbols). The x-axis represents SST temperatures, while the y-axis 
	 represents JWST temperatures. The error bars represent the 0.025 and 0.975 
	 quantiles of the posterior distributions of the parameters. The dashed 
	 line shows where the SST and JWST temperatures would be equal.
	 $T^\mathrm{cr}_\mathrm{sur}$ for the SST data of IP Tau is unconstrained.}\label{f05}
\end{figure}

\begin{figure}
			 \centering
			 \includegraphics[width=\linewidth]{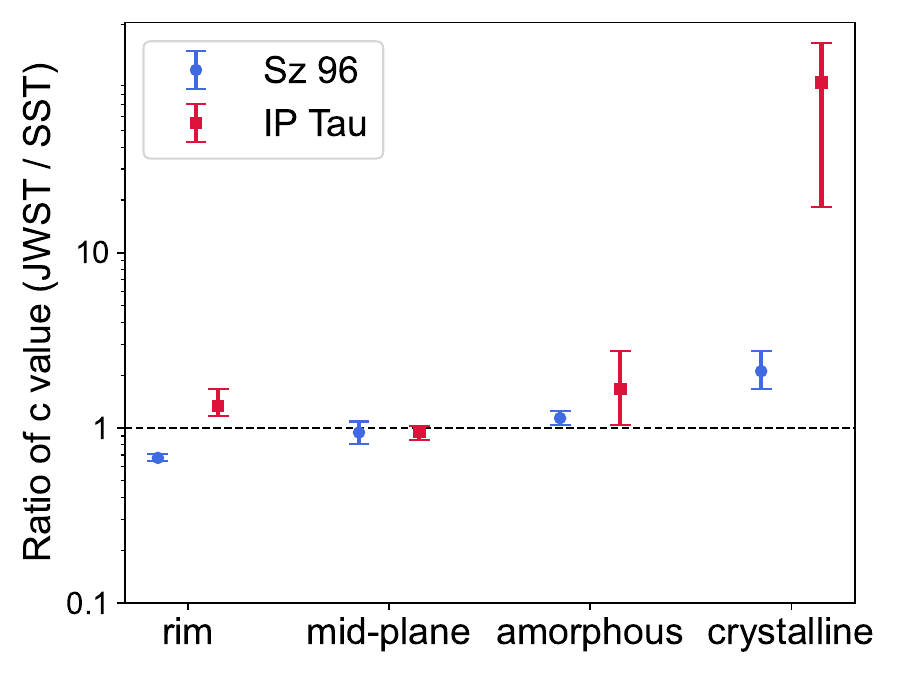}
	 \caption{The ratio of the $c$ value (JWST/SST) for Sz 96 (circles) 
	 and IP Tau (squares). The error bars represent the 0.025 and 
	 0.975 quantiles of the ratio, computed using samples drawn 
	 from the posterior distributions of the $c$ values for the 
	 JWST and SST data. The $c$ value for each component represents 
	 the solid angle as seen by the observer for the rim and mid-plane, 
	 and the dust mass in the surface region for amorphous and 
	 crystalline components.}\label{f06}
\end{figure}

The fitting results for Sz 96 and IP Tau are presented in 
Figs~\ref{f03} and \ref{f04}, respectively.
The medians of the posterior distributions of the fitting parameters, 
together with the RMS residuals of the best-fitting models for each spectrum, 
are summarised in \cref{t03}.

\begin{table*}
	\centering
    \caption{The medians and the 0.025 and 0.975 quantiles of the posterior 
			distributions of the parameters and the RMS residual for each spectrum.
			The parameter $c^\mathrm{am}_\mathrm{sur}$ 
			and $c^\mathrm{cr}_\mathrm{sur}$ represent the sum over each type of dust component.
			The units of $c$ parameters are $10^{-18}$~sr for the rim, $10^{-15}$~sr for the mid-plane,
			and $10^{-20}\mathrm{g \ cm^{-2} \ sr}$ for the surface components.}
    \label{t03}
    \begin{tabular}{l|cccccccccc}
        \hline
        Target & Data & $T_\mathrm{rim}$ (K) & $T_\mathrm{mid}$ (K)& $T^\mathrm{am}_\mathrm{sur}$ (K)
				& $T^\mathrm{cr}_\mathrm{sur}$ (K) & $c_\mathrm{rim}$ & $c_\mathrm{mid}$
				& $c^\mathrm{am}_\mathrm{sur}$ & $c^\mathrm{cr}_\mathrm{sur}$ 
				& Residual (RMS)\\
        \hline\hline
        \multirow{2}{*}{Sz 96}
				& SST & $984^{+17}_{-16}$ & $167^{+4}_{-4}$ & $411^{+10}_{-9}$ & $199^{+9}_{-8}$
				& $5.4^{+0.2}_{-0.2}$ & $1.2^{+0.2}_{-0.2}$ & $4.7^{+0.4}_{-0.4}$ & $17^{+5}_{-4}$
				&2.2\%\\
      	& JWST &  $1088^{+0.6}_{-0.6}$ & $163^{+0.2}_{-0.2}$ & $409^{+0.3}_{-0.3}$ & $176^{+0.2}_{-0.2}$ 
				& $3.6^{+0.01}_{-0.01}$ & $1.1^{+0.01}_{-0.01}$ & $5.3^{+0.02}_{-0.02}$ & $36^{+0.3}_{-0.3}$
				&2.0\%\\
				\hline
        \multirow{2}{*}{IP Tau}
				& SST &  $1281^{+135}_{-91}$ & $153^{+4}_{-5}$ & $458^{+67}_{-47}$ & $1150^{+333}_{-630}$ 
				& $3.2^{+0.4}_{-0.6}$ & $3.4^{+0.4}_{-0.3}$ & $3.0^{+0.2}_{-0.1}$ & $0.1^{+0.4}_{-0.04}$
				&3.3\%\\
				& JWST &  $1225^{+0.6}_{-0.6}$ & $156^{+0.06}_{-0.04}$ & $438^{+0.3}_{-0.3}$ & $168^{+0.8}_{-0.7}$ 
				& $4.2^{+0.01}_{-0.01}$ & $3.1^{+0.01}_{-0.01}$ & $5.0^{+0.01}_{-0.01}$ & $9.5^{+0.23}_{-0.19}$
				&1.6\%\\
        \hline
    \end{tabular}
\end{table*}

The top panels of \cref{f03} show that Sz 96 exhibits a decrease 
in the emission from the rim, whereas the mid-plane and surface 
components exhibit no significant variations in the JWST data 
compared with the SST data. In the middle panels, the observed features  
around 10~$\mathrm{\mu m}$ are dominated by amorphous  
silicates and are well reproduced by a combination of  
amorphous forsterite and enstatite with grain sizes of 0.1, 2, and 5~$\mathrm{\mu m}$.
The features around 20~$\mathrm{\mu m}$ are reproduced by the emission  
features of crystalline forsterite grains with sizes of  
0.1~$\mathrm{\mu m}$ and crystalline enstatite  
grains with a size of 5.0~$\mathrm{\mu m}$ in addition to  
the amorphous silicates. Despite small differences in the 
best-fitting grain sizes, we do not interpret these as 
meaningful changes in the grain size distribution between 
the two epochs, as the spectral features of each dust 
species included in the best-fitting model do not change 
significantly with grain size. 
The RMS residual is 2.2\% for the SST data and 2.0\% for the JWST data.

For IP Tau, the top panels of \cref{f04} show that the  
rim and the surface components become brighter, while the  
mid-plane component exhibits no significant change in 
the JWST data. For the SST data, $T^\mathrm{cr}_\mathrm{sur}$ is 
unconstrained (see Appendix \ref{app:b}). Since the posterior 
of $T^\mathrm{cr}_\mathrm{sur}$ is biased toward higher temperatures, 
we set $T^\mathrm{cr}_\mathrm{sur}=T_\mathrm{rim}=1281\,\mathrm{K}$ in 
the best-fitting model in \cref{f04} for practical purposes, although all subsequent discussions of 
the variability are based on the values listed in \cref{t03}. The SST data  
show that the features around 10~$\mathrm{\mu m}$ are
reproduced by the amorphous forsterite and enstatite  
grains with sizes of 0.1, 2, and 5~$\mathrm{\mu m}$,  
and the crystalline forsterite and enstatite with sizes  
of 0.1 and 5~$\mathrm{\mu m}$. In contrast, those 
in the JWST data are mainly reproduced by amorphous 
forsterite and enstatite with grain sizes of 0.1 
and 2~$\mathrm{\mu m}$. The features around 20~$\mathrm{\mu m}$ 
in the JWST data are reproduced by crystalline forsterite 
grains with sizes of 0.1 and 2~$\mathrm{\mu m}$ in addition to 
the amorphous silicates, as in Sz~96. 
The RMS residual is 3.3\% for the SST data and 1.6\% for the JWST data.

The temperatures of disc components for Sz 96  
and IP Tau are shown in \cref{t03} and \cref{f05}.  
For Sz~96, the rim temperature increases by 104\,K, 
whereas the surface crystalline component varies by 23\,K,
and the surface amorphous dust and mid-plane shows no 
significant change. 
In the case of IP Tau, no significant changes in
the temperature parameters are found in the JWST data compared to
the SST data except for $T^\mathrm{cr}_\mathrm{sur}$.
The mean of the posterior distribution for $T^\mathrm{cr}_\mathrm{sur}$ (1150\,K) in  
the SST data of IP Tau should not be interpreted as the best-fitting value.
The posterior distribution has a lower bound at 300\,K, with no samples 
found below this value.

Interestingly, in both the SST and JWST data of Sz 96, 
as well as in the JWST data of IP Tau, 
the temperature of the crystalline component    
is currently as low as 170--200\,K, and lower than that of the  
amorphous silicates (410--460\,K). Crystallisation is 
understood to occur through high-temperature 
annealing ($\gtrsim$ 1000\,K). If the lower temperature 
obtained in this analysis has always been representative 
of the crystalline component, it is consistent 
with the possibility that re-amorphisation operates in the inner disc and 
crystalline silicates are transported to larger radii.
This is discussed in further detail in \Cref{s5-2}.

The ratios of the $c$ values in the model of the JWST data  
to those of the SST data are shown in \cref{f06}. 
The $c$ values for the amorphous and crystalline dust 
represent the sum over each type of dust component.
The $c$ values for the rim and mid-plane represent the emitting
area of the rim and mid-plane, respectively, while the $c$ values 
for the surface amorphous and crystalline dust represent the dust 
mass in the surface region.
For Sz 96, the $c$ value for the rim decreases by 33\%, while  
that for the mid-plane does not change significantly
and the amorphous and crystalline dust in the surface increase by 13\%
and 112\%, respectively.
In IP Tau, the $c$ value for the mid-plane does not change significantly,
whereas those for the rim and the amorphous dust increase by 
31\% and 67\%, respectively. The $c$ value for the crystalline dust  
in the surface shows a significant increase, although  
with large uncertainty due to the unconstrained  
$T^\mathrm{cr}_\mathrm{sur}$.

For the JWST fits, the estimated parameter uncertainties $\approx 0.1$\%
are much smaller than the residuals $\approx 2$\%. This is because the parameter 
uncertainties reflect the local curvature of the $\chi^2$ surface around its minimum, 
rather than the overall goodness of fit. Consequently, they are primarily governed 
by the statistical uncertainties of the data. The statistical uncertainties of the 
JWST spectra, which include photon Poisson noise and detector read noise and correspond 
to a signal-to-noise ratio $\approx$1000, are responsible for the small 
parameter uncertainties. The larger residuals likely arise from systematic 
uncertainties and physical components not captured by the current model.

The results using the simple averaging method for crystalline silicates
optical properties (see \Cref{s3-2}) are presented in Appendix~\ref{app:a}.
The variabilities of the model parameters for IP~Tau are consistent. 
However, for Sz~96, the variability of the $T$ and $c$ parameters 
for the mid-plane and crystalline surface components 
is opposite to that presented in \Cref{s4}, as discussed in Appendix~\ref{app:a}. 
Therefore, we note that the interpretation of the $c$ variability 
of the mid-plane and crystalline surface components 
discussed in \Cref{s5-3} is not robust, as it may depend on the method 
used to combine the refractive index data for crystalline silicates.

\section{Discussion}\label{s5}
\subsection{The discrepancy between the best-fitting model 
and the observed data}\label{s5-1}
The current model successfully reproduces the overall  
shapes of the mid-infrared spectra, but a closer  
inspection reveals discrepancies between the model  
spectrum and the observed data, particularly for Sz 96, 
where more distinct spectral features are observed 
compared to IP Tau. In the case of JWST data  
of Sz 96, for example (\cref{f07}), the main  
discrepancies are as follows: (1) two broad bumps in the  
9--10~$\mathrm{\mu m}$ region in the observed data, which are
not captured by the model, (2) two peaks in the 
10--10.5~$\mathrm{\mu m}$ region in the model, which are not 
observed, and (3) a mismatch in the shoulder feature around 11  
and 18~$\mathrm{\mu m}$.

Regarding (1), these bumps could be reproduced by crystalline  
enstatite features. However, its contribution in the  
best-fitting model is limited to a level too small to account  
for the two bumps. While the features in the  
9--10~$\mathrm{\mu m}$ region could be reproduced by  
allowing a higher temperature for crystalline enstatite,  
we do not consider this possibility, because assuming  
different temperatures for crystalline forsterite and  
enstatite would be unrealistic. 

(2) and (3) may be attributed to shape effects  
used to calculate the dust mass absorption coefficients in  
the model. While we adopted the DHS approach in this  
study, \cite{J2024} used different methods, including  
Gaussian Random Field (GRF; \citealt{M2007}) method and  
laboratory aerosol measurements \citep{T2006}. These  
alternative models may reproduce the observed features  
but they may fail to reproduce the features beyond  
20~$\mathrm{\mu m}$ that are well matched by the DHS  
model used in this study (see \Cref{s5-2} and fig. 2 in  
\citealt{J2024}).

Although minor discrepancies remain as described above, 
no further tuning was performed, as the fits are very 
good and the mismatches result in only local and small 
residuals of less than 5\%.

\subsection{Temperature of crystalline silicate}\label{s5-2}

\begin{figure}
	\centering
	\includegraphics[width=\linewidth]{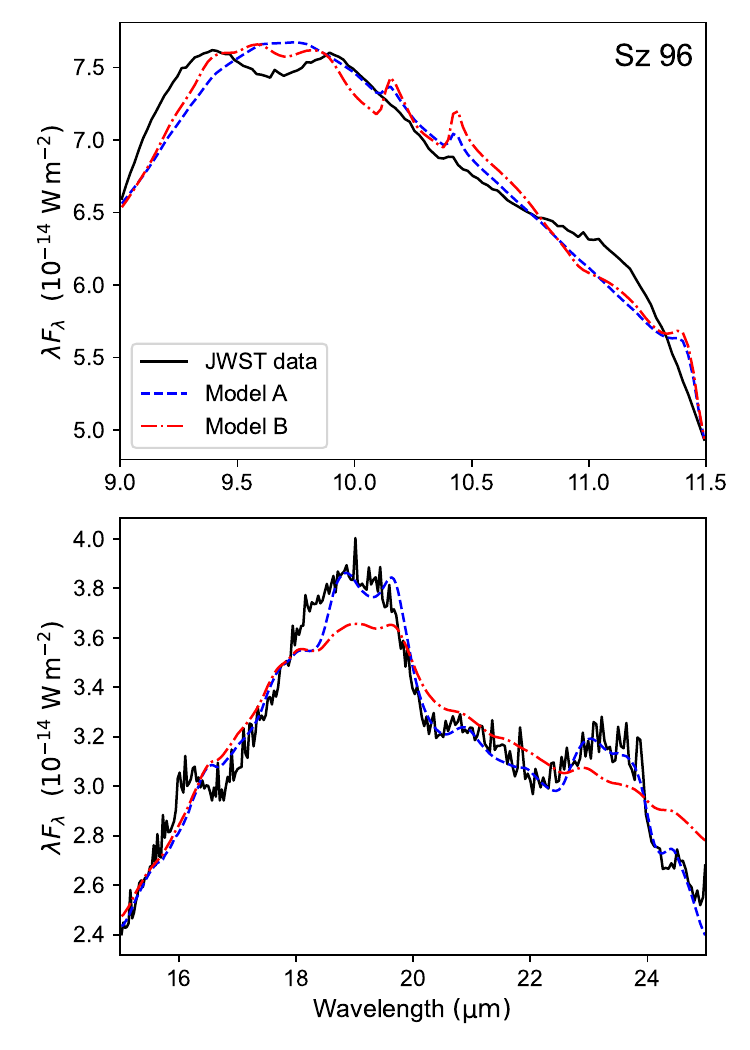}
	\caption{Best fit for JWST MIRI MRS data of Sz 96. Model A 
	(dashed curve) allows different temperatures for 
	amorphous and crystalline components, while their temperatures 
	are set to be the same in Model B (dash-dotted curve). 
	The top panel shows an enlarged plot around 10~$\mathrm{\mu m}$, 
	and the bottom panel shows an enlarged plot around 20~$\mathrm{\mu m}$.
	The best-fitting values of the $T$ parameters for model B are 
	$T_\mathrm{rim}=1089$\,K, $T_\mathrm{mid}=130$\,K, and 
	$T_\mathrm{sur}=406$\,K.}\label{f07}
\end{figure}

As discussed in \Cref{s4}, in the SST and JWST data of Sz  
96 and the JWST data of IP Tau, the crystalline silicates  
are found at lower temperatures than the amorphous  
component. This result arises from the attempt to  
reproduce the spectra, where crystalline features are  
prominent at $\sim$ 18~$\mathrm{\mu m}$ but weaker at  
$\sim$ 10~$\mathrm{\mu m}$. To verify the robustness of  
this result, we performed a fitting analysis on the JWST  
data of Sz 96 using a modified model in which the  
crystalline and amorphous components were constrained to  
have the same temperature. We define the original model  
as Model A and the modified model with the temperature  
constraint as Model B. The best-fitting spectra for each  
model are shown in \cref{f07}. The RMS residual  
is significantly larger for Model B than for  
Model A, with the values of 2.035(1)\% and 2.692(3)\%, respectively, 
where the values in parentheses indicate $3\sigma$ uncertainties. 
This trend becomes more  
evident at wavelengths longer than 20~$\mathrm{\mu m}$;  
2.579(3)\% for Model A and 4.055(6)\% for Model B. While  
the reproduced spectra around 10~$\mathrm{\mu m}$ are  
almost the same, the crystalline feature around  
20~$\mathrm{\mu m}$ is better reproduced by Model A than  
by Model B. We thus conclude that our finding of lower  
temperature of crystalline silicates than that of the  
amorphous component is robust.

Similar results have been reported in several studies  
using SST data of T Tauri stars. \cite{B2008} modelled  
the SST IRS data of T Tauri stars and confirmed that the  
crystalline fraction in the cooler components (130\,K)  
was 10--15 times higher than that in the warmer  
components (270\,K). \cite{O2009} analysed the spectra of  
108 T Tauri stars observed as part of the c2d Spitzer  
legacy programme \citep{E2003}, and found that  
crystalline features were detected 3.5 times more  
frequently in the wavelength range $\lambda > 20 \  
\mathrm{\mu m}$ compared to $\lambda < 10 \  
\mathrm{\mu m}$. However, \cite{O2010} proposed that in  
regions with a low crystalline fraction (i.e. the hotter  
components), crystalline features are buried in the  
amorphous features due to contrast issues. In the present  
study, using the JWST MIRI MRS data with higher spectral  
resolution and signal-to-noise ratio compared to the SST  
IRS data should make the crystalline features much  
clearer. Thus, our analysis of the JWST spectra suggests that  
the crystalline features are more prominent at longer  
wavelengths than at shorter wavelengths, which is not due  
to contrast issues.

The fact that crystals exist at lower temperatures  
suggests that they are located in the outer regions of  
the disc. However, crystalline silicates are generally  
believed to be localised in the innermost part of the  
disc or on the disc surface near the central star, where  
crystallisation efficiently occurs via annealing
\citep{J2024a}. The mechanism that leads to the presence  
of crystalline silicates in the outer region is under debate.  
\cite{G2004} suggested that material in the inner disc is heated and  
processed as it moves inward within the disc, and  
partially redistributed to the outer disc through  
turbulence and circulation.
\citet{G2019} showed that magnetocentrifugally driven 
disc winds can uplift submicron- to micron-sized 
crystalline grains from the inner disc surface and allow 
them to re-enter the disc at larger radii. Their model 
demonstrates that 1~$\mathrm{\mu m}$-sized crystalline grains at $\sim$ 0.1~au 
can be transported outward and re-enter the disc at distances 
of 1--10~au. \cite{J2024a} also proposed the following  
hypothesis: in the inner region of the disc, the  
crystallised silicates are easily re-amorphised by  
high-energy particles and cannot survive, thus  
crystalline silicates are found in the relatively outer,  
low-temperature region. Existing measurements show that Sz 96 
has a typical X-ray luminosity of $\sim 3.1\times10^{30} \ \mathrm{erg\,s^{-1}}$
\citep{G2006}, while no published X-ray luminosity is available for IP Tau. 
The re-amorphisation process by high-energy particles was 
quantitatively evaluated by \cite{G2009}. 
The relationship between crystalline fraction and X-ray luminosity, 
which can trace stellar activity and energetic ions from stellar winds, 
was investigated based on model fitting to SST/IRS data of T Tauri stars, 
but no statistically significant correlation was found \citep{O2010}. 
Identifying the mechanism requires detailed modelling of mixing due to disc winds and  
re-amorphisation. Nevertheless, the fact that we  
obtain consistent results with previous studies  
even using JWST data, which reveal crystalline features  
more clearly than before, highlights the robustness of  
this trend.

For the SST data of IP Tau, the posterior distribution of  
$T^\mathrm{cr}_\mathrm{sur}$ is broad and lacks a clear peak  
due to the low signal-to-noise ratio of the data, spanning from 
$\sim 300$\,K to the edge of the prior range at 1500\,K.
Even when the prior range is extended, the posterior distribution 
correspondingly broadens toward higher temperatures.
This may indicate the presence of 
high-temperature crystalline dust located in the 
inner disc of IP~Tau, although both the SST and JWST 
spectra around 10~$\mathrm{\mu m}$ do not show 
corresponding crystalline dust features. Given  
that the best-fitting model for the JWST data yields no  
contribution from crystalline enstatite components, we  
also performed the same analysis on the SST data with  
crystalline enstatite excluded. The result shows that the  
posterior distributions of $T_\mathrm{rim}$,  
$T_\mathrm{mid}$, $T^\mathrm{am}_\mathrm{sur}$, and  
$T^\mathrm{cr}_\mathrm{sur}$ have peaks at 1307, 157,  
408, and 38\,K, respectively, resulting in a lower  
temperature for the crystalline dust. The fit quality  
does not change significantly when crystalline enstatite  
is included: the mean absolute residual is 3.27(7)\% for  
the model without crystalline enstatite and 3.32(8)\% for  
the model including crystalline enstatite, where the values 
in parentheses indicate $3\sigma$ uncertainties. Thus, we  
caution that the temperature and mass of the crystalline  
dust in the SST data of IP Tau are highly uncertain and  
sensitive to the choice of crystalline species included  
in the model. Therefore, we exclude the variability of  
crystalline dust of IP Tau from the discussion in  
\Cref{s5-3}.

\subsection{Mechanism of Variability}\label{s5-3}

\begin{figure*}
	\centering
	\includegraphics[width=\linewidth]{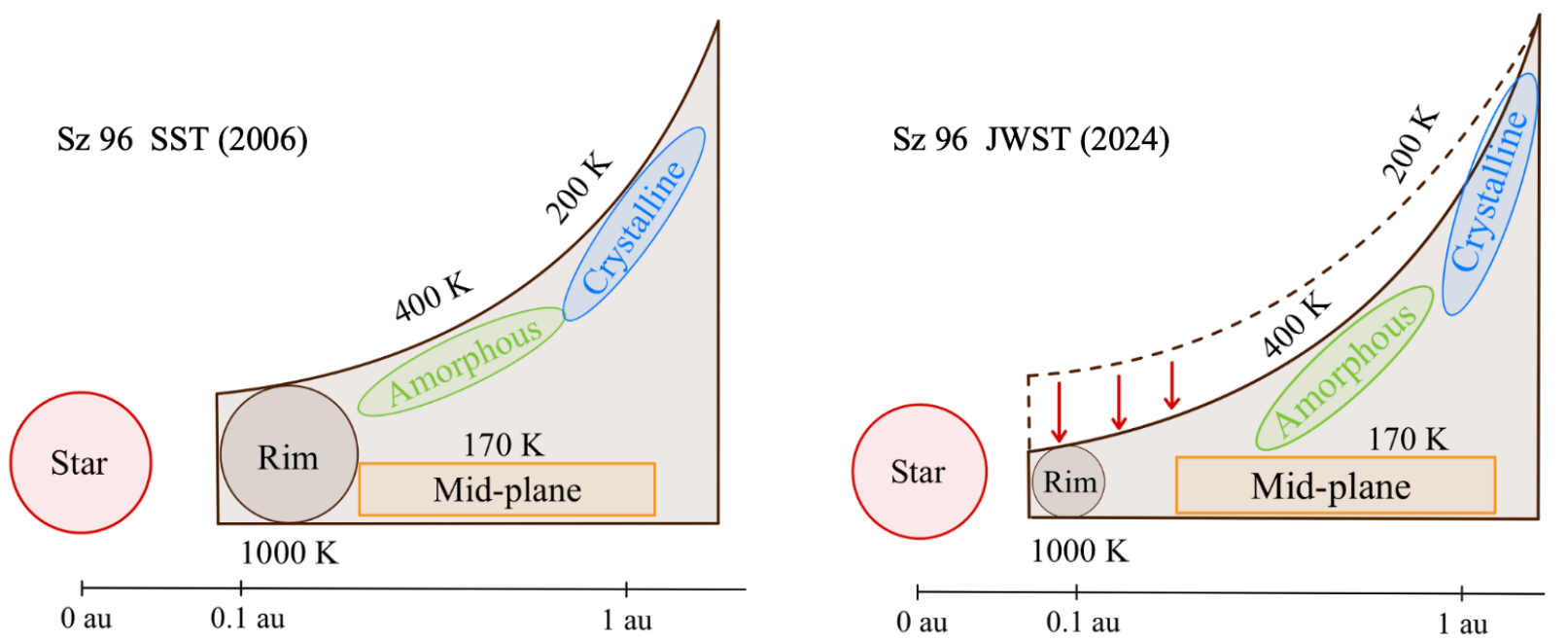}
	\caption{Changes in the disc structure which explain 
	the variability of Sz 96 between the SST (left) and JWST (right) observations.
	As the scale height of the inner region of the disc decreases, 
	the rim becomes smaller, and both the amorphous and 
	crystalline surface experience an increase in the solid angle 
	as seen from the central star, leading to an increase in 
	the input energy.}\label{f08}
\end{figure*}

\begin{figure*}
	\centering
	\includegraphics[width=\linewidth]{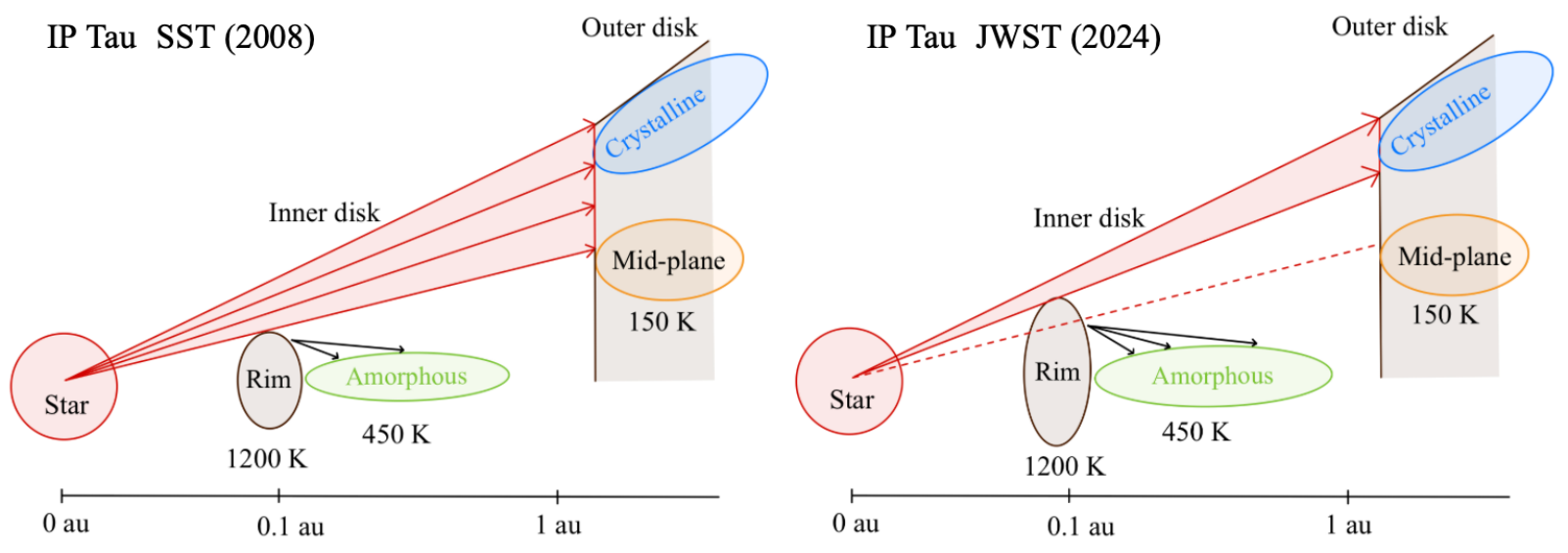}
	\caption{Changes in the disc structure which explain 
	the variability of IP Tau between the SST (left) and JWST (right) observations.
	As the height of the rim at the inner disc becomes larger, 
	the amorphous dust gains more input energy from the rim and
	the outer disc is shadowed \citep{E2011}.}\label{f09}
\end{figure*}

In the present study, the time variation of the spectra 
is analysed for two T Tauri stars with different disc 
structures, covering temporal baselines of 18 years for 
Sz 96 and 16 years for IP Tau. Here, we discuss the mechanism of the  
variability with a simple model. The reference model is  
the one proposed by \cite{E2011}. They explained the  
variability of a pre-transitional disc by changing the  
scale height of the inner disc. Based on this idea, we  
discuss the possibility that the observed spectral  
changes are caused by structural changes in the inner  
region of the discs. In addition, a disc structure with  
crystalline dust in the outer region is assumed  
(\Cref{s5-2}).

Sz 96 has a full disc, as suggested by its infrared excess. 
Based on our fitting results, the emitting  
area of the rim decreases, while that of the mid-plane
shows no significant change, and the mass of the
surface amorphous and crystalline dust increase.
This variability can be explained by a reduction in 
the disc scale height in the inner region. 
We illustrate the proposed scenario in detail in \cref{f08}.
We consider a disc structure in which the  
innermost region contains the rim, followed by a surface  
region with amorphous dust and the mid-plane that lies at  
the bottom of the optically thin layer. In the outer  
colder regions, a surface layer of crystalline dust is  
present. Although the rim temperatures in the SST and 
JWST models are similar, the scale height can differ 
because it is governed by the mid-plane temperature, which 
is not directly constrained.
Vertical mixing within the inner rim, for example induced 
by magnetically driven turbulence \citep{F2017}, 
can modify its internal vertical temperature structure and 
thereby lead to variations in the rim height. As the scale height 
of the inner disc becomes smaller, the vertical extent of the rim 
decreases and the rim subtends a smaller solid angle toward the observer.
At the same time, a reduced rim allows a larger fraction of stellar 
radiation to reach the disc surface.
Consequently, the surface layer containing amorphous and crystalline 
silicates intercepts more stellar energy, increasing the amount 
of dust that emits in the mid-infrared.

The disc around IP Tau is classified as a pre-transitional  
disc. Our analysis shows that the emitting area of the  
rim increases while that of the mid-plane does not change significantly, 
whereas the mass of the amorphous dust in the surface  
increases in the JWST data compared with the SST data. 
This variability can be explained by an increase in  
the scale height of the inner disc, which casts a larger  
shadow over the outer disc (\citealt{E2011}; see  
\cref{f09}). The assumed disc structure consists of the  
inner disc rim, a gap with optically thin amorphous dust,  
and the outer disc with a surface layer of cold  
crystalline dust and the mid-plane at its base. This  
structure was proposed by \cite{E2011} to explain the  
strong 10~$\mathrm{\mu m}$ silicate emission feature of  
pre-transitional discs including IP Tau. The inner disc
rim expands its emitting area from the SST era to the JWST era.
As the rim becomes larger, the amorphous dust that mainly receives 
radiation from the rim gains more input energy, increasing the mass 
of dust emitting in the mid-infrared.
Simultaneously, the larger shadow cast on the outer disc 
can reduce the mid-plane emitting area, but because the mid-plane 
constitutes the base of the disc, the decrease is slight and not significant.

These scenarios explain all of the observed temporal variations as resulting from 
changes in the disc structure. However, the observed changes in the dust 
mass may alternatively arise from dust clouds lifted into the disc atmosphere via 
turbulence, and this possibility cannot be excluded.
Although the model used in this study successfully reproduces 
the mid-infrared spectra with a small number of parameters, it is based 
on a simple spectral decomposition and does not explicitly account for 
the detailed disc structure.
Therefore, the structural changes inferred from variations in the model 
parameters involve inherent ambiguities.
The next step in uncovering the full nature of the dust disc is to 
construct models that incorporate the overall disc geometry and are 
consistent with the full spectral energy distribution.

\section{Conclusions}\label{s6}
We analysed mid-infrared spectra of two T~Tauri stars, Sz~96 and IP~Tau, 
obtained by SST and JWST. The spectra were fitted by a model that separates 
the emission into optically thin and thick components. 
The optical properties of crystalline silicates were primarily 
calculated using the Bruggeman rule, while the simple averaging 
method was also tested for comparison.
We quantitatively evaluated the variability of the dust 
composition and temperature distribution of the discs over 
approximately 20~years between the SST and JWST observations. 
From these results, we derived the following three conclusions.

\begin{enumerate}
	  \item In the disc around Sz 96, the emitting area of the rim   
		decreases, while that of the mid-plane does not change significantly.
		The dust mass in the surface increases for both amorphous and crystalline components.
		However, we found that the variability of the mid-plane and 
		crystalline dust become opposite when using the simple averaging method.
		In the disc around IP Tau, the emitting  
		area of the rim increases while that of the mid-plane  
		does not change significantly, whereas the amorphous dust mass increases. 
		The variability of the crystalline  
		dust in the surface of IP Tau cannot be reliably  
		determined, as it strongly depends on the dust species  
		included in the model. These variabilities for IP~Tau are consistent 
		regardless of the method used to calculate the optical properties of crystalline silicates.
		\item The observed spectral variations may be qualitatively  
		explained by the change in the scale height of the inner  
		region of the discs. The temporal variation is different  
		in Sz 96 and IP Tau, which may also be explained by  
		their different disc structures. Further studies with  
		self-consistent models of disc structure and radiation  
		transfer are desirable.
		The longer time baseline observation would help determine 
		whether the observed rim size variations are part of transient 
		events or indicative of secular structural evolution.
		Future JWST monitoring over several years would provide 
		valuable constraints on the time-scales and physical drivers of 
		these structural changes.
    \item The temperature of crystalline silicates is  
    lower than that of amorphous silicates in the SST and  
		JWST data of Sz 96 and the JWST data of IP Tau. Although  
		it was suggested in the SST observations of some discs,  
		the higher spectral resolution and sensitivity of JWST  
		robustly establish this finding. This suggests that crystalline  
		silicates are depleted in the hot inner disc, which  
		could be due to re-amorphisation in the inner disc or  
		transportation to outer radii by disc winds.
\end{enumerate}

\section*{Acknowledgements}
T.M. acknowledges the support by KAKENHI grant number 23K20237.
Y.A. acknowledges the support by KAKENHI grant number 24K00674 and 20H05847.
T.J.H. acknowledges a Dorothy Hodgkin Fellowship, UKRI guaranteed 
funding for a Horizon Europe ERC consolidator grant (EP/Y024710/1) 
and UKRI/STFC grant ST/X000931/1.
C.J.L. acknowledges the support by NASA through the NASA Hubble 
Fellowship grant No. HST-HF2-51535.001-A awarded by the Space 
Telescope Science Institute, which is operated by the Association 
of Universities for Research in Astronomy, Inc., for NASA, under 
contract NAS5-26555.
N.P.B. acknowledges support from NSF grant No. AST-2205698 and NASA/Space 
Telescope Science Institute grants JWST-GO-03271 and JWST-GO-05460.
C.G-R. acknowledges support from ANID-Millennium Science Initiative 
Program-Center Code NCN2024 001.
V.V.G. acknowledges support from ANID-Millennium Science Initiative Program-Center 
Code NCN2024\_001, CATA-Basal FB210003 and from FONDECYT Regular 1221352.
We acknowledge Klaus M. Pontoppidan for his reduction of the JWST data, 
and Gianni Cataldi for his valuable guidance 
on MCMC sampling.
This research has made use of the NASA/IPAC Infrared Science Archive, 
which is operated by the Jet Propulsion Laboratory, California Institute 
of Technology, under contract with the National Aeronautics and Space Administration.
This work is based on observations made with the Spitzer Space Telescope, 
which was operated by the Jet Propulsion Laboratory, California Institute 
of Technology under a contract with NASA.
This work is based on observations made with the NASA/ESA/CSA James Webb 
Space Telescope. The data were obtained from the Mikulski Archive for Space 
Telescopes at the Space Telescope Science Institute, which is operated by the 
Association of Universities for Research in Astronomy, Inc., under NASA contract 
NAS 5-03127 for JWST. These observations are associated with programme \#50403.
The authors declare no conflicts of interest.

\section*{Data Availability}
The JWST MIRI MRS data used in this study are accessible  
from the Mikulski Archive for Space Telescopes  
(MAST\footnote{\url{https://archive.stsci.edu/missions-and-data/jwst}})  
as part of the programme 3228 (PI: L. Ilsedore Cleeves). The SST  
IRS data for Sz 96 and IP Tau were retrieved from the  
NASA/IPAC Infrared Science Archive  
(IRSA\footnote{\url{https://irsa.ipac.caltech.edu/frontpage/}}) as part  
of the IRS Enhanced Spectrophotometric Products. Sz 96  
was observed under programme 179 (PI: Neal Evans), and IP  
Tau under programme 50403 (PI: Nuria Calvet).



\bibliographystyle{mnras}
\bibliography{M-lab}



\appendix

\section{The results using the simple averaging method for 
crystalline silicates optical properties}\label{app:a}

\begin{table*}
	\centering
    \caption{The medians and the 0.025 and 0.975 quantiles of the posterior 
			distributions of the parameters and the RMS residual for each spectrum.
			The parameter $c^\mathrm{am}_\mathrm{sur}$ 
			and $c^\mathrm{cr}_\mathrm{sur}$ represent the sum over each type of dust component.
			The units of $c$ parameters are $10^{-18}$~sr for the rim, $10^{-15}$~sr for the mid-plane,
			and $10^{-20}\mathrm{g \ cm^{-2} \ sr}$ for the surface components.}
    \label{tb1}
    \begin{tabular}{l|cccccccccc}
        \hline
        Target & Data & $T_\mathrm{rim}$ (K) & $T_\mathrm{mid}$ (K)& $T^\mathrm{am}_\mathrm{sur}$ (K)
				& $T^\mathrm{cr}_\mathrm{sur}$ (K) & $c_\mathrm{rim}$ & $c_\mathrm{mid}$
				& $c^\mathrm{am}_\mathrm{sur}$ & $c^\mathrm{cr}_\mathrm{sur}$ 
				& Residual (RMS)\\
        \hline\hline
        \multirow{2}{*}{Sz 96}
				& SST & $984^{+17}_{-16}$ & $163^{+3}_{-2}$ & $440^{+15}_{-14}$ & $213^{+11}_{-11}$
				& $5.4^{+0.2}_{-0.2}$ & $1.5^{+0.1}_{-0.1}$ & $3.7^{+0.5}_{-0.4}$ & $12^{+3}_{-2}$
				&2.1\%\\
      	& JWST &  $1079^{+0.4}_{-0.4}$ & $130^{+0.1}_{-0.1}$ & $415^{+0.3}_{-0.3}$ & $360^{+1.0}_{-0.9}$ 
				& $3.7^{+0.01}_{-0.01}$ & $4.7^{+0.01}_{-0.01}$ & $6.0^{+0.02}_{-0.02}$ & $1.3^{+0.01}_{-0.01}$
				&2.5\%\\
				\hline
        \multirow{2}{*}{IP Tau}
				& SST &  $1304^{+96}_{-85}$ & $154^{+3}_{-4}$ & $463^{+51}_{-38}$ & $1031^{+449}_{-584}$ 
				& $3.1^{+0.5}_{-0.4}$ & $3.3^{+0.3}_{-0.2}$ & $3.0^{+1.3}_{-0.9}$ & $0.02^{+0.24}_{-0.01}$
				&3.4\%\\
				& JWST &  $1224^{+0.5}_{-0.6}$ & $154^{+0.1}_{-0.1}$ & $442^{+0.3}_{-0.3}$ & $188^{+0.5}_{-0.6}$ 
				& $4.2^{+0.01}_{-0.01}$ & $3.5^{+0.01}_{-0.01}$ & $4.9^{+0.01}_{-0.01}$ & $4.2^{+0.06}_{-0.06}$
				&1.6\%\\
        \hline
    \end{tabular}
\end{table*}

\begin{figure*}
	\centering
	\includegraphics[width=\linewidth]{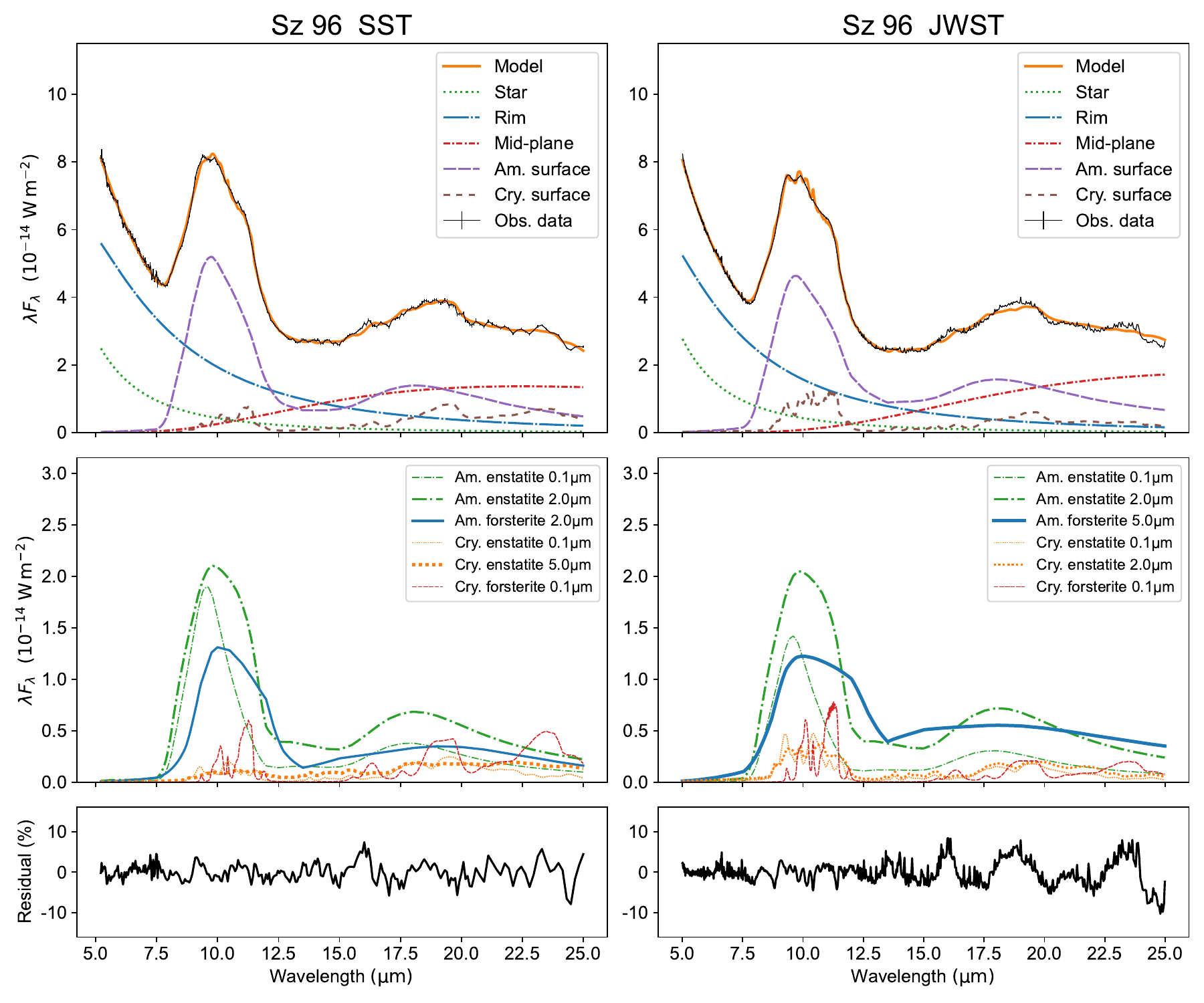}
	\caption{Fitting results of Sz 96 using the simple averaging method 
for crystalline silicates optical properties: overall fitting results 
	(top), contributions from each dust component (middle), and 
	residual of the best-fitting model (bottom) for SST (left) and 
	JWST (right). In the middle panel, the emissions from the star, 
	rim, and mid-plane, as well as components contributing less than 
	0.5\% to the total flux, are not shown.}\label{fa1}
\end{figure*}

\begin{figure*}
	\centering
	\includegraphics[width=\linewidth]{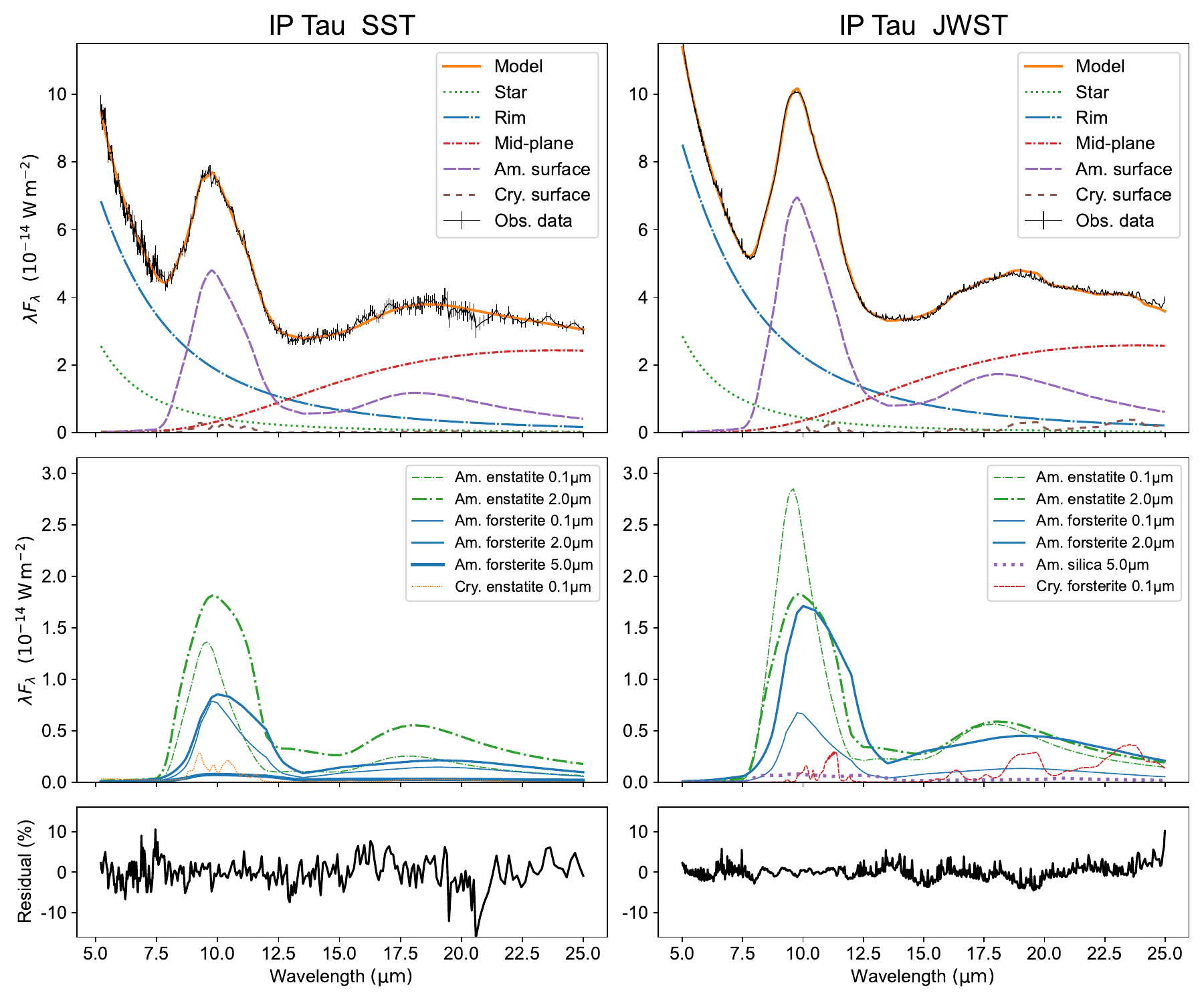}
	\caption{Fitting results of IP Tau using the simple averaging method 
for crystalline silicates optical properties: overall fitting results 
	(top), contributions from each dust component (middle), and 
	residual of the best-fitting model (bottom) for SST (left) and 
	JWST (right). In the middle panel, the emissions from the star, 
	rim, and mid-plane, as well as components contributing less than 
	0.5\% to the total flux, are not shown.}\label{fa2}
\end{figure*}

\begin{figure}
			 \centering
			 \includegraphics[width=\linewidth]{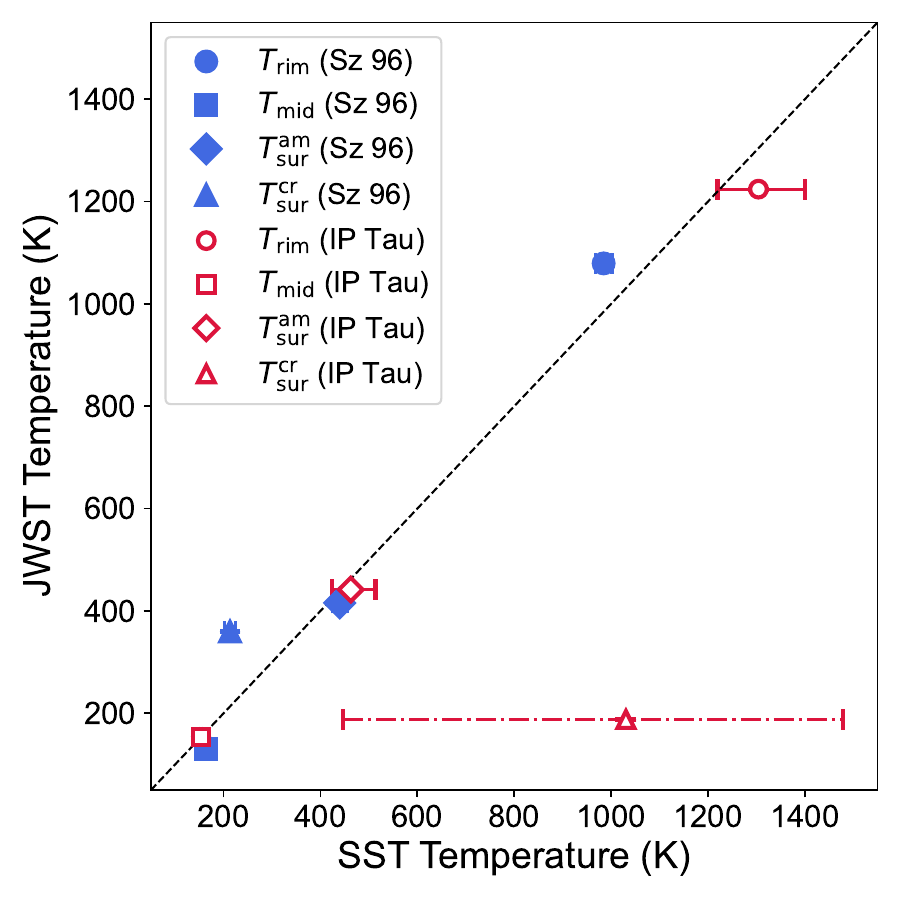}
	 \caption{The temperature parameters for Sz 96 (filled symbols) and IP Tau (open symbols) 
	 using the simple averaging method for crystalline silicates 
	 optical properties. The x-axis represents SST temperatures, while the y-axis 
	 represents JWST temperatures. The error bars represent the 0.025 and 0.975 
	 quantiles of the posterior distributions of the parameters. The dashed 
	 line shows where the SST and JWST temperatures would be equal.
	 $T^\mathrm{cr}_\mathrm{sur}$ for the SST data of IP Tau is unconstrained.}\label{fa3}
\end{figure}

\begin{figure}
			 \centering
			 \includegraphics[width=\linewidth]{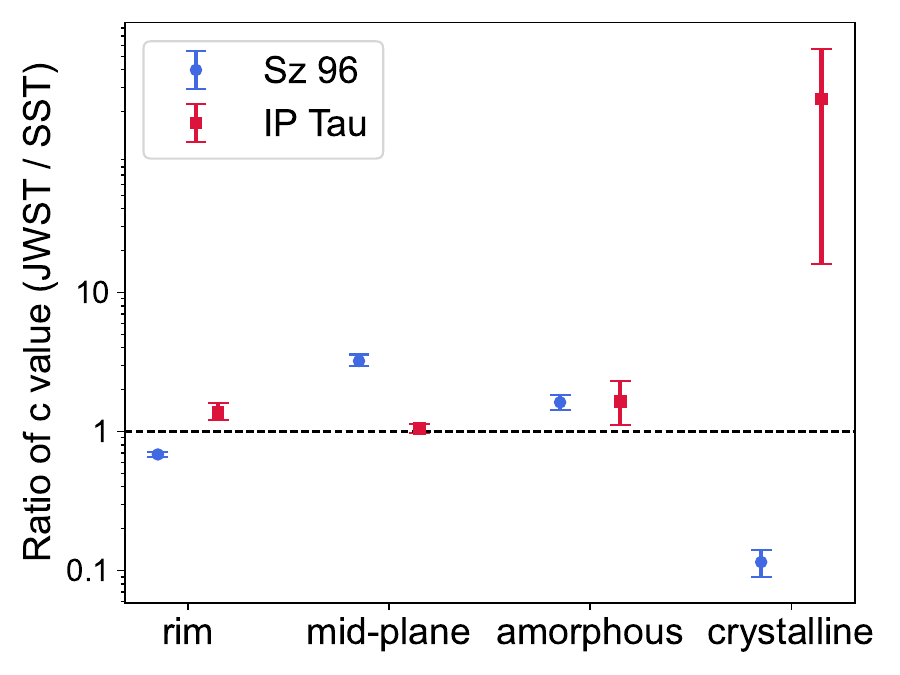}
	 \caption{The ratio of the $c$ value (JWST/SST) for Sz 96 (circles) 
	 and IP Tau (squares) using the simple averaging method for 
	 crystalline silicates optical properties. The error bars represent the 0.025 and 
	 0.975 quantiles of the ratio, computed using samples drawn 
	 from the posterior distributions of the $c$ values for the 
	 JWST and SST data. The $c$ value for each component represents 
	 the solid angle as seen by the observer for the rim and mid-plane, 
	 and the dust mass in the surface region for amorphous and 
	 crystalline components.}\label{fa4}
\end{figure}

We performed the same analysis using the simple averaging method 
for crystalline silicates optical properties (see \Cref{s3-2}).
The fitting results for Sz 96 and IP Tau are presented in Figs \ref{fa1} and \ref{fa2}, 
respectively. The best-fit parameters are summarised in \cref{tb1}.
The temperature and $c$ parameters are shown in Figs \ref{fa3} and
\ref{fa4}, respectively.

For IP~Tau, the trends in the variability of the parameters 
remain consistent with those obtained using the Bruggeman rule in \Cref{s4}. 
However, for Sz~96, the mid-plane temperature decreases by 33\,K, 
whereas that of the crystalline dust component increases by 147\,K, 
which are different compared to the variabilities presented in \Cref{s4}. 
Consequently, the $c$ parameter for the mid-plane increases by 213\%, 
while that for the crystalline dust decreases by 89\%, 
both exhibiting variability trends different from those shown in \Cref{s4}.

Therefore, we note that the interpretation of the variability for IP~Tau 
remains robust, whereas that for the mid-plane and crystalline surface dust 
in Sz~96 likely depends on the method used to calculate the optical properties 
of crystalline silicates.

\section{The posterior distributions}\label{app:b}

\begin{figure*}
    \centering
    \includegraphics[width=\linewidth]{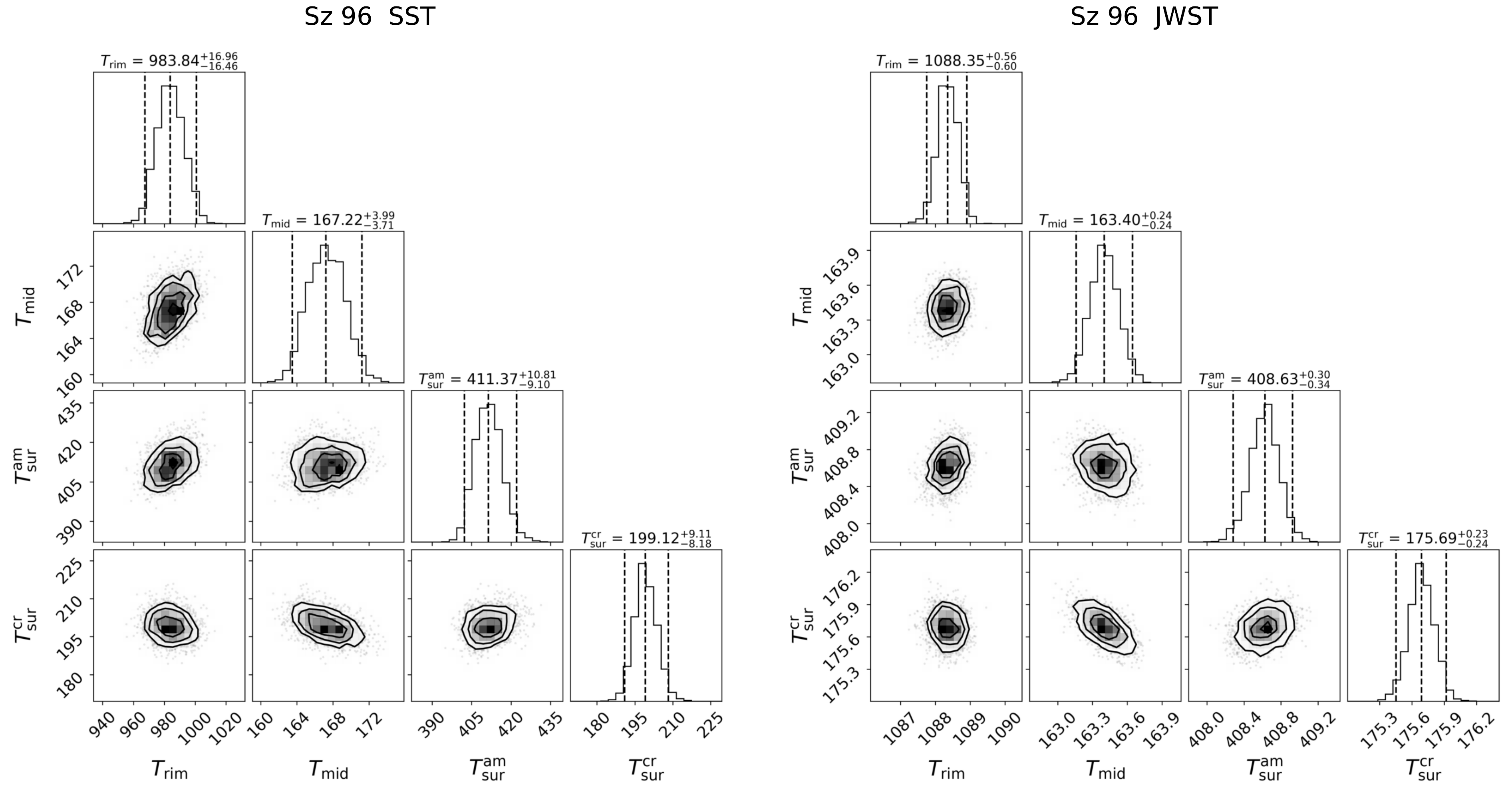}
    \caption{The posterior distributions of the temperature parameters
		for the SST (left) and JWST (right) data of Sz 96.
		The dashed lines correspond to the quantiles $q=0.025,0.5,0.975$.}\label{fb1}
\end{figure*}

\begin{figure*}
    \centering
    \includegraphics[width=\linewidth]{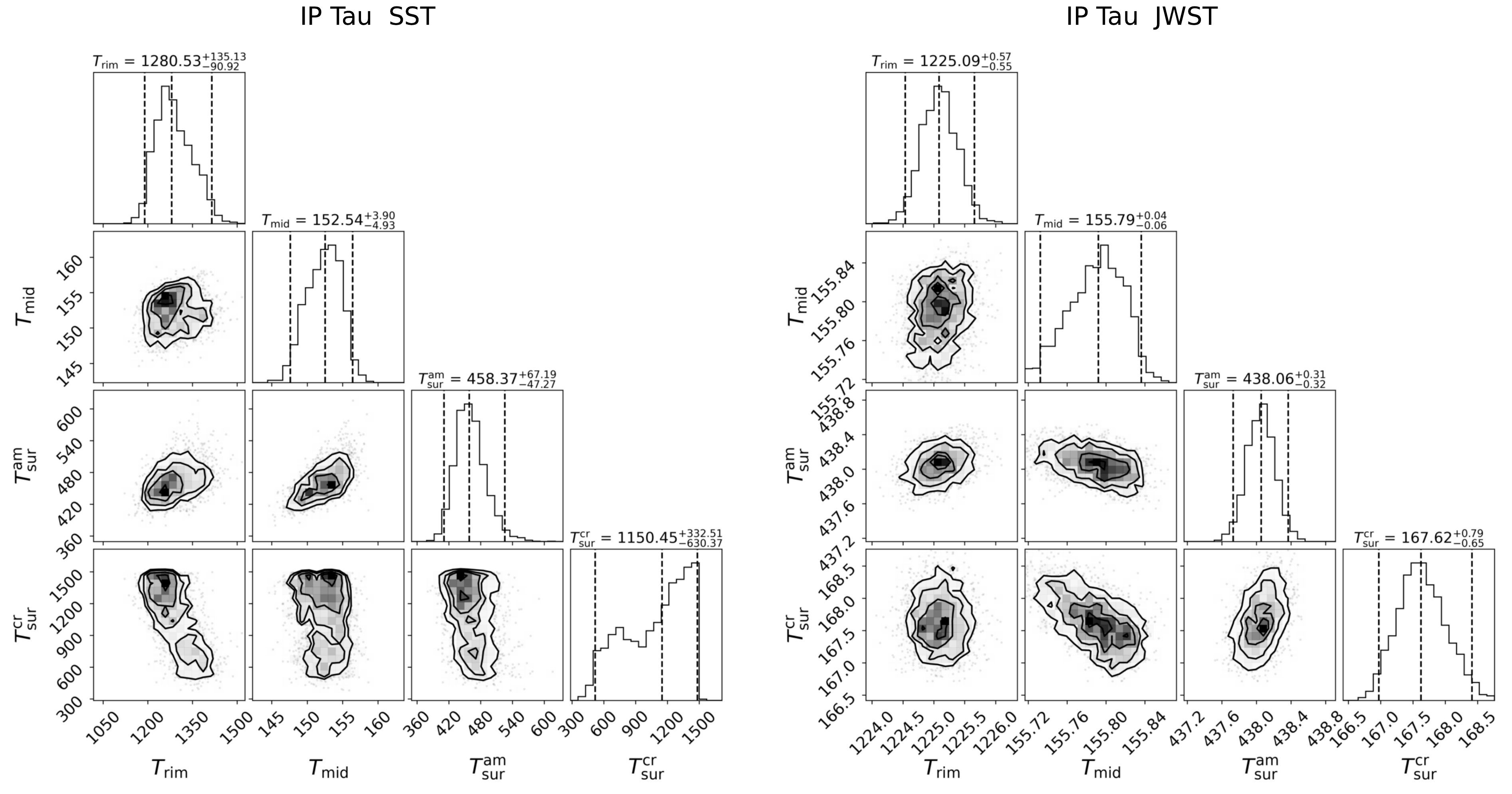}
    \caption{The posterior distributions of the temperature parameters
		for the SST (left) and JWST (right) data of IP Tau.
		The dashed lines correspond to the quantiles $q=0.025,0.5,0.975$.
		For the SST data of IP Tau, the posterior distribution of 
		$T^\mathrm{cr}_\mathrm{sur}$ is broad and extends to the upper limit of 
		the prior range, indicating that this parameter is unconstrained.
    This is the only case among all parameters and datasets where the posterior 
		distribution reaches the prior boundary.}\label{fb2}
\end{figure*}

The posterior distributions of the temperature parameters  
are shown in \cref{fb1} for Sz 96 and \cref{fb2} for IP  
Tau. These distributions are those obtained using the Bruggeman method. 
Most parameters exhibit clear peaks, except for  
$T^\mathrm{cr}_\mathrm{sur}$ in the SST data of IP Tau,  
whose posterior distribution spans from a lower limit of  
approximately 300\,K to an upper limit corresponding to  
the dust sublimation temperature of about 1500\,K. Within  
this range $T^\mathrm{cr}_\mathrm{sur}$ significantly  
affects only $c^\mathrm{cr}_\mathrm{sur}$. In the  
best-fitting model shown in the left panel of \cref{f04}, we  
set $T^\mathrm{cr}_\mathrm{sur}$ equal to be the  
best-fitting value of $T_\mathrm{rim}$.

\bsp	
\label{lastpage}
\end{document}